\documentclass[journal]{IEEEtran}
\usepackage{amsmath,amsfonts}
\usepackage{array}
\usepackage[caption=false,font=normalsize,labelfont=sf,textfont=sf]{subfig}
\usepackage{textcomp}
\usepackage{stfloats}
\usepackage{url}
\usepackage{verbatim}
\usepackage{graphicx}
\usepackage{cite}
\usepackage{amsmath}
\usepackage{amssymb}
\usepackage{graphicx}
\usepackage{xcolor}
\definecolor{darkgreen}{rgb}{0.0, 0.39, 0.0}
\usepackage{enumitem}
\usepackage{url}
\usepackage{pifont}
\usepackage[linesnumbered,ruled,vlined]{algorithm2e}
\usepackage{booktabs}
\usepackage{multirow}
\usepackage{colortbl}
\usepackage{hyperref}
\usepackage{cleveref}
\usepackage{alltt}

\hypersetup{
    colorlinks=true,
    linkcolor=blue,      
    citecolor=blue,      
    urlcolor=black       
}

\begin{document}
\bstctlcite{IEEEexample:BSTcontrol}
\title{Privacy-Preserving LLM Embedding Transmission\\for End-Cloud Collaboration}

\author{
Shuaifan Jin, Xiaoyi Pang,  Zhibo Wang,~\IEEEmembership{Senior Member, IEEE,} He Wang, Jiacheng Du,\\Jiahui Hu, Kui Ren,~\IEEEmembership{Fellow, IEEE}
\thanks{© 2026 IEEE.  Personal use of this material is permitted.  Permission from IEEE must be obtained for all other uses, in any current or future media, including reprinting/republishing this material for advertising or promotional purposes, creating new collective works, for resale or redistribution to servers or lists, or reuse of any copyrighted component of this work in other works.}
\thanks{This work was supported by National Natural Science Foundation of China under Grant U24B20182, Scientific Research Innovation Capability Support Project for Young Faculty (SRICSPYF-ZY2025016), the R\&D Program of Zhejiang (Grants No. 2026SDXT014, 2026C02A1246), Shandong Provincial Natural Science Foundation (Grant No. ZR2023LZH018), Ganpo Talent Support Program (Grant No. GPJC20260071), Jiangxi Provincial Natural Science Foundation under Grant No. 20262BAC200470, and Youth Talent Training Program for Early-career Professionals under Grant No. 20262BEJ730057. (Corresponding author: Zhibo Wang.)}
\thanks{Shuaifan Jin, Zhibo Wang, He Wang, Jiacheng Du, Kui Ren are with the State Key Laboratory of Blockchain and Data Security, Zhejiang University, Hangzhou 310027, China, and the College of Computer Science and Technology, Zhejiang University, Hangzhou 310027, China, and also with Hangzhou High-Tech Zone (Binjiang) Institute of Blockchain and Data Security, Hangzhou, Zhejiang 310012, China (e-mail: shuaifanjin@zju.edu.cn; zhibowang@zju.edu.cn; wanghe\_71@zju.edu.cn; jiahuihu@ncu.edu.cn; jcdu@zju.edu.cn; kuiren@zju.edu.cn).}
\thanks{Xiaoyi Pang is with Hong Kong University of Science and Technology (e-mail: xypang@whu.edu.cn).}
\thanks{Jiahui Hu is with the School of Artificial Intelligence, Nanchang University, 330031, China (e-mail: jiahuihu@ncu.edu.cn).}
}




\maketitle

\begin{abstract}
Recent studies improve on-device language model (LM) inference through end-cloud collaboration, where the end device retrieves useful information from cloud databases to enhance local processing, known as Retrieval-Augmented Generation (RAG). Typically, to retrieve information from the cloud while safeguarding privacy, the end device transforms original data into embeddings with a local embedding model. However, the recently emerging Embedding Inversion Attacks (EIAs) can still recover the original data from text embeddings (e.g., training a recovery model to map embeddings back to original texts), posing a significant threat to user privacy. To address this risk, we propose EntroGuard, an entropy-driven perturbation-based embedding privacy protection method, which can protect the privacy of text embeddings while maintaining retrieval accuracy during the end-cloud collaboration. Specifically, to defeat various EIAs, we perturb the embeddings to increase the entropy of the recovered text in the common structure of transformer-based recovery models, thus steering the embeddings toward meaningless texts rather than original sensitive texts during the recovery process. To maintain retrieval performance in the cloud, we constrain the perturbations within a bound, applying the strategy of reducing them where redundant and increasing them where sparse. Moreover, EntroGuard can be directly integrated into end devices without requiring any modifications to the embedding model. Extensive experimental results demonstrate that EntroGuard effectively reduces privacy leakage metrics to near-zero levels against learning-based EIAs while also mitigating optimization-based EIAs with negligible loss of retrieval performance.
\end{abstract}


\section{Introduction}

\IEEEPARstart{I}{n} recent years, driven by advancements in large language models (LLMs) and the proliferation of diverse end devices, on-device deployment of LLMs has become a significant trend, which provides personalized services while protecting user privacy. However, end-side LLMs typically have a smaller parameter scale and limited capabilities, making them prone to generating unreliable information during reasoning~\cite{zhu2024understanding}. A primary strategy for correcting such hallucinations is to integrate external factual knowledge into the user's original input through End-Cloud Collaboration, namely retrieval-augmented generation (RAG). Specifically, end devices employ dedicated embedding models to convert raw sensitive text input into human-unrecognizable embeddings~\cite{wang2024crafter} and subsequently transmit them to the cloud to retrieve corresponding knowledge from the cloud database to augment local reasoning. Such embeddings enable knowledge retrieval without disclosing original data.

However, recent research~\cite{morris2023text,li2023sentence,morris2023language,wan2024information,chen2024text} proposed Embedding Inversion Attacks (EIA), demonstrating that embeddings still pose privacy risks since they can be used to recover users' original sensitive text input. Such attacks can be categorized into two types, i.e., optimization-based attacks and learning-based attacks. In optimization-based attacks~\cite{morris2023text,chen2024text}, a random text is initialized and then iteratively optimized to make its corresponding embedding closely match the target embedding. On the other hand, learning-based attacks~\cite{li2023sentence,wan2024information} involve training a recovery model, i.e., a generative model, to map embeddings back to their original text. This type of attack is more threatening than the previous one, as attackers often leverage powerful transformer-based large language models with strong generative capabilities for recovery. Hence, embeddings transformed from raw sensitive inputs are not as secure as previously assumed, and the continuous transmission of these payloads in end-cloud systems exposes users to severe privacy risks.

Currently, securing these distributed transmissions presents a practical networking dilemma. While standard network-layer protocols like TLS prevent in-transit eavesdropping, they leave the plaintext embeddings fully exposed to an honest-but-curious cloud provider. Conversely, applying cryptographic protocols like homomorphic encryption or secure multi-party computation introduces severe bandwidth expansion and latency overhead, rendering them impractical for real-time collaborative inference over resource-constrained edge-to-cloud networks. Meanwhile, there is limited discussion on defenses against Embedding Inversion Attacks in LLMs~\cite{li2024privlm}. Some works~\cite{liu2024mitigating,wan2024information,zeng2024privacyrestore} protect embeddings by transforming them into another form, such as an encrypted domain and frequency domain. However, these approaches require additional cloud-side computation to reverse the process so that the embeddings can be recognized by the cloud database. Therefore, they are not applicable in real-world scenarios since the third-party cloud is hard to be compromised to complete the reversal. Other defense works~\cite{zhou2023textobfuscator,morris2023text} achieve embedding protection by adding noise into embeddings, but we find that they can only defend against optimization-based EIAs while failing to defend against more powerful learning-based EIAs.

In this paper, we aim to safeguard end-user privacy against various EIAs while achieving precise external knowledge retrieval without additional cloud-side computation. To achieve this, we need to overcome two main challenges that are mutually constrained. The first challenge is \textit{how to defend against different EIAs universally, especially learning-based EIAs}. Due to the fundamental differences in attack mechanisms, it is hard to defend against both types of attacks in a universal way. Moreover, learning-based EIAs may use various recovery models, which further increases the difficulty of successfully protecting embeddings. The second challenge is \textit{how to maintain the retrieval performance while effectively protecting the privacy of embeddings}. Protecting embeddings inevitably alters them. However, excessive changes hinder precise information retrieval from the cloud database, while insufficient changes fail to safeguard privacy. In this context, it is challenging to balance privacy protection and retrieval performance.

To address the above challenges, we propose a plug-in application-layer embedding privacy protection method for end-cloud collaborative LLM reasoning, called EntroGuard. It can protect text embeddings against various EIAs and maintain their retrieval performance without additional cooperation from the cloud side. To this end, we first perturb the original text embedding to defeat the most powerful learning-based EIAs. Usually, learning-based EIAs adopt a variety of transformer-based generative models to map embeddings back to original texts, and transformer blocks are the common structure in such models. To achieve robust defense against different attack models, we propose an entropy-driven perturbation mechanism that perturbs the embedding to increase the entropy of the recovered information of the transformer blocks. This can effectively redirect the embedding away from the original sensitive text and steer it toward meaningless content throughout the recovery process. Then, to ensure that the disturbed embedding can still accurately retrieve the required external knowledge from the cloud database, we propose a bound-aware perturbation adaptation mechanism. It imposes strict constraints on the perturbation intensity to prevent the perturbation from being redundant or insufficient, thus retaining essential semantic information in embeddings while effectively securing its privacy.

Besides, existing metrics may not adequately reflect text privacy leakage in certain cases as they primarily focus on comparing similarity on text level. To evaluate privacy leakage from the perspective of semantic similarity, we leverage the concept of bidirectional entailment in semantic entropy and propose a semantic entailment rate metric, called BiNLI, as a complement to existing metrics.

Our main contributions can be summarized as follows:

\begin{itemize}[leftmargin=*]

\item We propose a novel privacy-preserving transmission approach, EntroGuard, which can preserve retrieval performance without requiring additional cooperation from the cloud side and effectively protect text privacy against representative EIAs, especially learning-based EIAs instantiated by different Transformer-based recovery models.

\item We propose Entropy-based Perturbation Generation and Bound-aware Perturbation Adaptation mechanisms, which serve as plug-in modules that can be seamlessly integrated into on-device black-box embedding models, safeguarding individuals' privacy prior to network transmission with zero bandwidth expansion and acceptable overhead.

\item  We introduce the BiNLI metric to facilitate a more comprehensive quantification of text privacy leakage. Extensive experiments demonstrate that EntroGuard outperforms the existing embedding protection methods in terms of superior privacy protection performance with negligible retrieval performance loss.

\end{itemize}

\section{Related work}
This section outlines the end-cloud collaboration paradigm in \cref{sec:Related0}, discusses privacy attacks in LLMs with a focus on recent embedding inversion attacks in \cref{sec:Related1}, and reviews existing protection schemes and their vulnerabilities in \cref{sec:Related2}.

\subsection{End-Cloud Collaboration} \label{sec:Related0}
End-cloud collaboration is becoming a prevailing paradigm, especially with the growing popularity of large language models, as it enables on-device applications to leverage the vast computational power and knowledge of the cloud while preserving users' original data~\cite{rahdari2025survey}. In former years, Kang et al.~\cite{kang2017neurosurgeon} proposed Neurosurgeon, a foundational End-Cloud Collaboration approach that determines the best layer to partition a DNN with the goal of reducing overall latency or energy usage. Jeong et al.~\cite{jeong2018computation} utilized split computing as a privacy-preserving approach to protect user-side data. Subsequently, Borzunov et al.~\cite{borzunov2022petals} demonstrated a decentralized, collaborative inference model, allowing a client to run large models by organizing multiple independent servers into a pipeline-parallel chain. More recently, this collaborative paradigm has been utilized to mitigate factual hallucination, an issue that becomes especially pronounced in on-device large language models given their limited scale. To counter this, a prevailing strategy is to employ a Retrieval-Augmented Generation (RAG) architecture~\cite{gao2023retrieval,ma2023query,lin2023ra}. It allows the on-device model to retrieve timely and factual knowledge from the cloud to ground its responses, effectively mitigating hallucinations~\cite{huang2024survey}. However, the necessary data transfer in any end-cloud collaboration system creates a potential privacy vulnerability, and recent studies have shown that this issue persists in large language models.

\subsection{Privacy attacks in LLMs} \label{sec:Related1}

In the early years, there existed several works~\cite{song2020information,pan2020privacy,fowl2022decepticons} proposed for the recovery of specific private information. Song et al.~\cite{song2020information} implemented an attribute inference attack in language models to obtain sensitive attributes in user input. Pan et al.~\cite{pan2020privacy} implemented pattern reconstruction attacks with fixed-pattern inputs, and keyword inference attacks, similar to attribute inference attacks. Although these attacks can obtain some sensitive information, their capabilities are limited by strict scenario requirements as they rely on specific datasets or data formats, such as genomics and ID numbers. Besides, in federated learning, Fowl et al.~\cite{fowl2022decepticons} extracted token and position embeddings to retrieve high-fidelity text by deploying malicious parameter vectors, but this type of attack is not within the scope of our consideration in this article due to different scenarios.

Recently, a new type of attack in large language models has emerged, i.e., embedding inversion attacks~\cite{morris2023text,chen2024text,morris2023language,li2023sentence,wan2024information}, aiming to revert original sensitive text inputs of users. Such attacks can be categorized into optimization-based EIAs and learning-based EIAs. In optimization-based EIAs, Morris et al.~\cite{morris2023text} generated text that closely aligns with a fixed point in the latent space when re-embedded. Building on this, Chen et al.~\cite{chen2024text} extended such attacks to languages other than English. Meanwhile, in learning-based EIAs, Li et al.~\cite{li2023sentence} reconstructed input sequences solely from their sentence embeddings via the powerful generation capability of existing LLMs. Morris et al.~\cite{morris2023language} reconstructed the input text based on the next-token probabilities of the language model. Wan et al.~\cite{wan2024information} also proposed a method to reconstruct the text input from embeddings in deeper layers. In summary, compared to previous privacy attacks that can only infer specific words, embedding inversion attacks can recover the whole sensitive text inputs of users, posing a much greater privacy threat to text embeddings in end-cloud collaboration.

\subsection{Privacy protection methods in LLMs} \label{sec:Related2}
For LLMs, several works have been proposed to protect text privacy, which can be broadly divided into two types, i.e., transformation-based and perturbation-based methods.

Transformation-based method always transforms the embedding to another domain to prevent attacks, e.g., frequency domain, where the protected embedding and the raw embedding vary a lot. Liu et al.~\cite{liu2024mitigating} use projection networks and text mutual information optimization to safeguard embeddings, but they need another projection network to maintain utility. Wan et al.~\cite{wan2024information} transform embedding into frequency domain, which also means that the database in the cloud must also be converted into the frequency domain format accordingly. These approaches rely on additional cooperation from clouds beyond retrieval, and become impractical when additional reversal processes cannot be executed in the third-party clouds or the reverse process goes wrong, as the transformed embeddings no longer align with those stored in the database.

Perturbation-based methods usually protect text privacy by differential privacy. Xu et al.~\cite{xu2020differentially} added an elliptical noise to the embedding space to balance privacy and utility in word replacement. Li et al.~\cite{li2023privacy} employed text-to-text privatization by differential privacy in token embeddings to privatize users' data locally. Du et al.~\cite{du2023dp} perturbed embedding matrices or hidden representations in the forward pass of LMs for differentially private fine-tuning and inference. Tong et al.~\cite{tong2025inferdpt} perturbed textual tokens in prompts or documents for privacy-preserving closed-box LLM generation. These methods share the perturbation-based paradigm, but they do not directly target text embeddings transmitted during end-cloud collaboration. Meanwhile, some perturbation-based methods focus on altering the original meaning of the text to protect privacy. Zhou et al.~\cite{zhou2022textfusion} hid private words through unrecognized words. Further, Zhou et al.~\cite{zhou2023textobfuscator} added random perturbations to clustered word representation for privacy. However, neither of the above two types of perturbation-based approaches can effectively defend against emerging EIAs and ensure cloud retrieval performance at the same time. Besides, there also exist several methods that add adversarial perturbations~\cite{goodfellow2014explaining, madry2017towards} to protect privacy. However, adversarial samples in text~\cite{goyal2023survey} are usually generated through modifications like insertion, deletion, swapping, or paraphrasing. In such situations, the adversarial text usually retains its original meaning, so privacy is still compromised when the text is recovered by EIAs.

In summary, previous protective methods either required the cloud to execute additional computation before retrieval tasks or lacked strong defense capabilities when facing embedding inversion attacks.

\section{Preliminary}

In this section, we first present the basic concepts of embedding models used on the end device in \cref{sec:Preliminary1}, then outline the fundamental components of retrieval‑augmented generation in \cref{sec:Preliminary1.5}.

\subsection{Embedding Model} \label{sec:Preliminary1}

The embedding model is a type of model that projects input text from its original format into a numerical embedding, such as a vector or matrix~\cite{patil2023survey,mao2016training}. It is often used to retrieve similar knowledge, as it aims to retain semantic and relational information of different texts in numerical space.

In the earlier years, one of the representative works of the embedding model is Word2vec~\cite{mikolov2013efficient} that developed the Continuous Bag-of-Words model and the Continuous Skip-gram model to transform words into vector representations, also referred to as word embeddings. In recent years, the prevailing powerful embedding model typically follows the dual encoder paradigm~\cite{ni2022large,ni2022sentence,gao2021simcse} in which queries and documents are encoded separately into a shared fixed-dimensional embedding space~\cite{ni2022large} to get a better retrieval performance. They usually adopt transformer architecture and use contrastive learning with in-batch sampled softmax loss:
\begin{equation}
\begin{split}
\mathcal{L}=\frac{e^{\operatorname{cos}\left(q_{i}, p_{i}^{+}\right) / \tau}}{\sum_{j \in \mathcal{B}} e^{\operatorname{cos}\left(q_{i}, p_{j}^{+}\right) / \tau}},
\end{split}
\label{eq:contrastive Loss}
\end{equation}
\noindent where $\operatorname{cos}(\cdot)$ denotes cosine similarity, $\mathcal{B}$ represents a mini-batch of examples, and $\tau$ is the softmax temperature, $q_{i}$ represents the queries, and $p_{i}^{+}$ represents the documents that can be factual knowledge.

\subsection{Retrieval-augmented generation} \label{sec:Preliminary1.5}
Retrieval-augmented generation (RAG) aims to utilize external information to address hallucinations and outdated knowledge~\cite{abootorabi2025ask}. Fundamentally, RAG enhances an LLM by first retrieving pertinent, up-to-date documents from an external knowledge base to ground the generation process in verifiable facts, rather than relying solely on the model's static training data. The primary implementation of retrieval-augmented generation follows four steps:

\begin{itemize}

\item Encoding: The entire user query is transformed into an embedding vector via embedding models.

\item Retrieval: This vector retrieves the top‑k most relevant documents from a database.

\item Augmentation: Retrieved content is merged with the original query.

\item Generation: The augmented input is fed into an LLM to produce the final response.

\end{itemize}

Notably, in typical RAG systems, retrieved documents are often much longer than the user's query, with documents spanning hundreds or even thousands of tokens, whereas queries typically contain only a few dozen.
Specifically, the user query represents a direct and concentrated expression of an individual's private intent, context, and personal details, while the retrieved documents are sourced from a vast and public corpus of general information. Therefore, within the scope of this work, we primarily focus on the protection of the user query, as it represents a potent source of privacy leaks.                                                                                                                                                                                                                            

\section{Problem Definition}
In this section, we first introduce a typical end–cloud collaboration system based on RAG in \cref{sec:Preliminary2}, then discuss the threat models that widely exist in reality in \cref{sec:Preliminary3}, and finally define the core problem addressed in this paper in \cref{sec:Preliminary4}. The key notations utilized throughout this paper are summarized in \cref{tab:notations}.

\begin{figure}[!t]
\centering
\includegraphics[width=0.8\columnwidth]{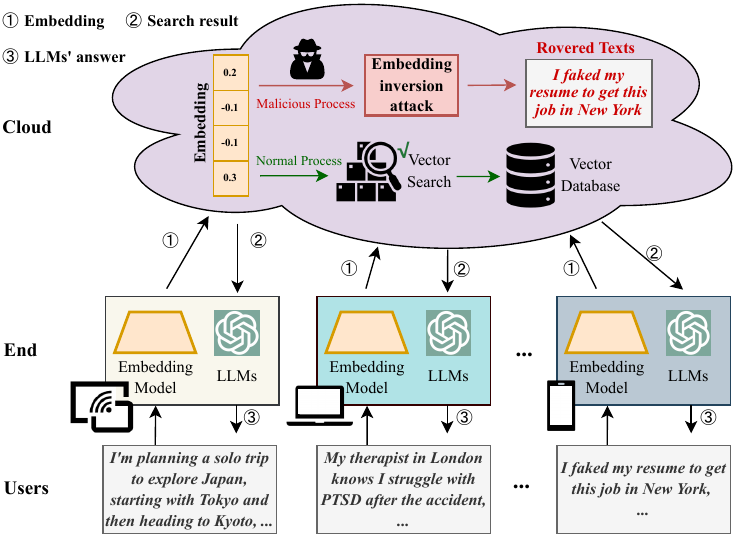}
\caption{System model. The cloud can infer private information from embeddings sent by end devices for knowledge retrieval.}
\label{fig:intro_pipe}
\end{figure}

\subsection{System model} \label{sec:Preliminary2}

In this paper, we consider a RAG-based end-cloud collaboration system. In this architecture, end devices with computationally limited models collaborate with a cloud server hosting a vast external knowledge base. The primary goal of this collaboration is to retrieve relevant external knowledge from the cloud, which helps to mitigate hallucinations and enhance the quality of their locally generated responses.

Specifically, the system operates in two phases. During the service initialization phase, the end device is equipped with an embedding model capable of projecting text into the same vector space as that used by the cloud's database. The inference phase, as illustrated in \cref{fig:intro_pipe}, then unfolds in several steps: First, the end device encodes the user's private input into an embedding vector. This vector is then transmitted to the cloud, which performs a similarity search to retrieve the top-k most relevant documents from its external knowledge base. Finally, these retrieved documents are transmitted back to the end device. The local LLM then integrates this external knowledge with the original user query to generate a more reliable and factually grounded response. 

Although transmitting embeddings instead of raw data has been generally considered an effective approach to prevent direct privacy leakage in end-cloud collaboration, recent research~\cite{morris2023text,chen2024text,morris2023language,li2023sentence,wan2024information} reveals that these embeddings may still be inverted to reconstruct sensitive user information, posing a significant privacy risk.

\begin{table}[!t]
\centering
\caption{Key Variable Notations}
\label{tab:notations}
\resizebox{0.98\columnwidth}{!}{
\begin{tabular}{cl}
\toprule
\textbf{Symbol} & \textbf{Description} \\
\midrule
$f$ & End-side embedding model. \\
$\Phi$ & Attacker's generative model (decoder). \\
$G_\theta$ & Parameterized perturbation generator model. \\
$S$ & Surrogate attacker model for training $G_\theta$. \\
$W_{dec}$ & Decoding matrix of a generative model. \\
$LN_s$ & Final normalization layer of surrogate model $S$. \\
$b_k$ & Output vector of the $k^{\text{th}}$ transformer block. \\
$\text{Distr}(k)$ & Word probability distribution from the $k^{\text{th}}$ block. \\
$\mathcal{D}$ & User's textual data distribution. \\
$\mathcal{X}$ & Text corpus for training. \\
$\mathcal{V}$ & Vocabulary of the generative model. \\
$x$ & Original user text. \\
$\hat{x}$ & Text recovered by an attacker. \\
$\mathbf{e}, \mathbf{e^0}$ & Raw (original) embedding vector of text $x$. \\
$\mathbf{e'}$ & Perturbed embedding vector from generator $G_\theta$. \\
$\mathbf{e''}$ & Final adapted embedding vector after bound adjustment. \\
$\epsilon$ & Perturbation bound (similarity threshold). \\
$\lambda$ & Scaling factor for perturbation adaptation. \\
\bottomrule
\end{tabular}}
\end{table}

\subsection{Threat model} \label{sec:Preliminary3}

\paragraph{Attack Scenarios} In reality, cloud providers may be incentivized to analyze users' personal data for commercial gains, including targeted advertising based on sensitive attributes (e.g., addresses or dietary preferences). This real-world scenario leads us to define our adversary as an honest-but-curious cloud. Rather than actively modifying data or execution workflows to compromise privacy, such an adversary strictly complies with all prescribed service protocols. However, it attempts to passively learn sensitive information by analyzing the legitimately received payloads, i.e., the uploaded embeddings. Since the cloud does not break any rules or tamper with data, this passive exploitation is highly stealthy and difficult for users to detect, posing a serious threat to privacy.

\paragraph{Adversaries' knowledge} In this scenario, the cloud provider acts as an adversary, directly threatening the privacy of user embeddings. This adversary is considerably powerful and is positioned to perform EIAs by leveraging the data it accesses. Specifically, it possesses the key prerequisites for such attacks:

\begin{itemize}[leftmargin=*]
    \item \textbf{Knowledge of the embedding model}: The adversary knows the user-side embedding model $f$, as the end device must use a model compatible with the cloud's vector database to enable retrieval.
    
    \item \textbf{Access to the user's embedding}: The adversary has direct access to the embedding $f(x)$ uploaded by the user, as the cloud performs similarity search directly on this query embedding.
\end{itemize}

\paragraph{Adversaries' strategy.} According to the recent emerging EIAs~\cite{morris2023text,li2023sentence,wan2024information,chen2024text}, the adversaries can recover the original sensitive user text $x$ from embeddings $f(x)$ mainly through the following two approaches. 

The first approach is optimization-based attack methods. It aims to retrieve the text $\hat{x}$ whose embedding $f(\hat{x})$ is maximally similar to the ground truth of original embedding, i.e., $\mathbf{e}=f(x)$. The iterative process can be formalized as an optimization problem, where the search for $\hat{x}$ is guided by the embedding model $f$, i.e., $\hat{x}=\arg \max _{x} \cos (f(x), e).$ Specifically, the text hypothesis for the next iteration will be obtained from the text hypothesis for the current iteration:
\begin{equation}
\begin{split}
p\left(x^{(t+1)} \mid e\right) & =\sum_{x^{(t)}} p\left(x^{(t)} \mid e\right) p\left(x^{(t+1)} \mid e, x^{(t)}, \hat{e}^{(t)}\right),
\end{split}
\label{eq:op attack Loss2}
\end{equation}
\noindent where the $e$ at the iteration $t$ is the embedding of the $x$ at the iteration $t$, i.e., $\hat{e}^{(t)}=f\left(x^{(t)}\right)$.

The second method is to implement a learning-based attack approach that aims to train a generative model $\Phi$ to reverse the mapping of the embedding model $f$, i.e., $\Phi(f(x)) \approx f^{-1}(f(x))=x.$ Generally, it mainly entails two steps. First, the attacker extracts text-embedding pairs $(x,f(x))$ from the corpus $\mathcal{X} = \left\{x_{1}, \ldots, x_{n}\right\}$ via the embedding model $f$ in the end device. Then these text-embedding pairs are utilized to train the generative model $\Phi$ by decreasing the cross-entropy between original and recovered texts:
\begin{equation}
\begin{split}
L_{CE}\left(x ; \theta_{\Phi}\right)=-\sum_{i=1}^{u} \log \left(\operatorname{P}\left(w_{i} \mid f(x), w_{0}, w_{1}, \ldots, w_{i-1}\right)\right),
\end{split}
\label{eq:learning attack Loss}
\end{equation}
\noindent where $x=[w_0,w_1,\dots,w_{u-1}]$ represents a sentence of length $u$. Finally, the trained generative model can be utilized to recover the users' sensitive input from the embeddings sent by the end device, i.e., $\hat{x}=\Phi(f(x))$.

\begin{table}[t]
\caption{Recovery and retrieval performance under different levels of Gaussian noise.} 
\label{table:noise_influence} 
\centering
\resizebox{0.8\columnwidth}{!}{
\begin{tabular}{cccc}
\toprule
\begin{tabular}[c]{@{}c@{}}\textbf{Noise}\\\textbf{level}\end{tabular} & \begin{tabular}[c]{@{}c@{}}\textbf{Optimization-}\\\textbf{based EIAs}\end{tabular} & \begin{tabular}[c]{@{}c@{}}\textbf{Learning-}\\\textbf{based EIAs}\end{tabular} & \textbf{Retrieval} \\
\midrule
0 & \ding{51} & \ding{51} & \ding{51} \\
$10^{-3}$ & \ding{51} & \ding{51} & \ding{51} \\
$10^{-2}$ & \ding{55} & \ding{51} & \ding{51} \\
$10^{-1}$ & \ding{55} & \ding{55} & \ding{55} \\
\bottomrule
\end{tabular}}
\end{table}

\begin{figure*}[!t]
    \centering
    \includegraphics[width=1.8\columnwidth]{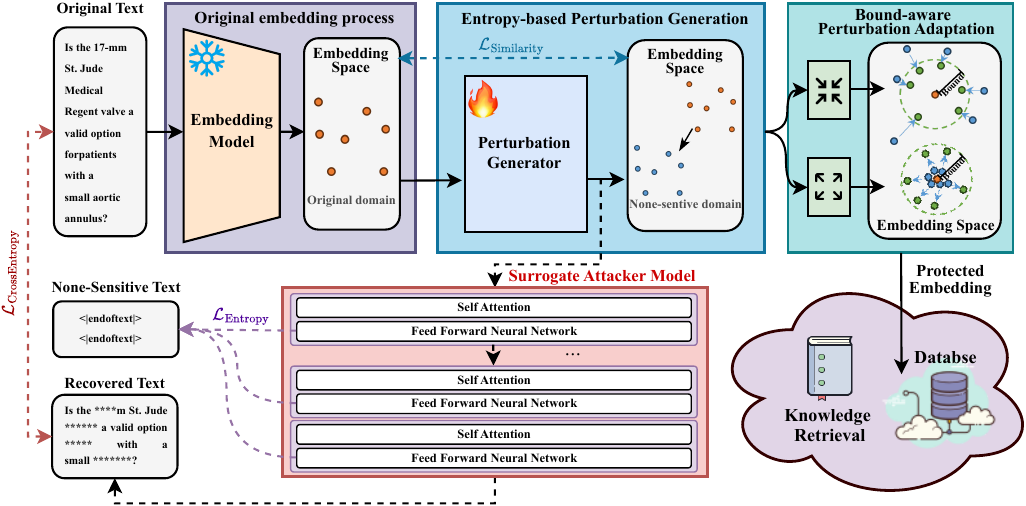}
    \caption{Pipeline of EntroGuard where the dashed arrows indicate the training process and the realization arrows indicate the inference process. During the training phase, a surrogate attacker model was built for optimizing perturbation generator in Entropy-based Perturbation Generation. In the inference phase, the original text is converted into an embedding through the embedding model and then processed by EntroGuard, including Entropy-based Perturbation Generation, Bound-aware Perturbation Adaptation, resulting in a protected embedding.}
    \label{fig:pipe}
\end{figure*}

Notably, observation from Morris et al.~\cite{morris2023text} showed that introducing a certain level of noise can help defend against optimization-based EIAs. However, we find that this approach is ineffective against learning-based EIAs via pre-experiments shown in \cref{table:noise_influence}. Considering that learning-based EIAs not only exhibit broader applicability but also demonstrate superior attack capabilities, we prioritize the prevention of learning-based attacks in this paper.

\subsection{Design Goals}
\label{sec:Preliminary4}

According to the system and threat models described, the inherent privacy risks in the RAG-based end-cloud architecture create a pressing need for a protection mechanism that allows users to perform accurate queries without leaking sensitive information. A significant challenge is that users have no control over the cloud provider's infrastructure or internal processes. This constraint, for instance, prevents the adoption of bespoke encryption schemes that would require the cloud to implement a corresponding decryption module. Moreover, even if such a deployment were possible, RAG typically relies on plaintext embeddings for similarity searches to maintain practical efficiency, leaving user privacy still exposed to the cloud provider.

Motivated by these challenges and constraints, our objective is to design a user-side protection mechanism that balances the competing demands of privacy and utility. This mechanism, denoted by a parameterized function $G_\theta$, where $\theta$ represents its trainable parameters, transforms an original embedding into a protected one. Our design is guided by two primary goals.

\paragraph{Privacy Preservation} 
The foremost goal is to protect user data from embedding inversion attacks. The mechanism must ensure that any text reconstructed from a protected embedding is semantically dissimilar to the original input. We formalize this by aiming to minimize the similarity between the original text $x$ and the text reconstructed by the adversary's model $\Phi$ from the protected embedding $G_\theta(f(x))$:
\begin{equation}
\underset{\theta}{\text{minimize}} \quad \mathbb{E}_{x\sim\mathcal{D}}\big[\text{Sim}\big(x, \Phi(G_\theta(f(x)))\big)\big],
\end{equation}
where $\text{Sim}(\cdot, \cdot)$ denotes the text similarity metric, and the expectation $\mathbb{E}$ is taken over the user's private textual data distribution $\mathcal{D}$. In our experiments, we approximate the expectations over $\mathcal{D}$ using the training splits of the target datasets (e.g., PersonaChat or MS MARCO).

\paragraph{Utility Preservation} 
While ensuring privacy, the protection must not compromise the core functionality of the RAG system. The documents retrieved using the protected embedding should be almost as relevant as those retrieved with the original one. We express this as a constraint on the Top-K recall, where $k$ is the number of retrieved documents, requiring it to remain above a predefined tolerance threshold $1 - \delta_{\text{ret}}$:
\begin{equation}
\mathbb{E}_{x \sim \mathcal{D}}\Big[\frac{|\mathcal{R}_k(f(x)) \,\cap\, \mathcal{R}_k(G_\theta(f(x)))|}{k}\Big] \ge 1 - \delta_{\text{ret}},
\end{equation}
where $\mathcal{R}_k(\cdot)$ denotes the set of top-$k$ retrieved documents, and the fraction calculates the recall by measuring the overlap between the sets retrieved for the original embedding $f(x)$ and the protected one $G_\theta(f(x))$.

\paragraph{Overall Problem Formulation} 
Integrating these two goals, the central challenge of this paper is to find the optimal parameters $\theta$ for our protection mechanism $G_\theta$. This is captured by the following constrained optimization problem:
\begin{equation}
\begin{aligned}
\min_{\theta} \quad & \mathbb{E}_{x\sim\mathcal{D}}\big[\text{Sim}\big(x, \Phi(G_\theta(f(x)))\big)\big] \\
\text{subject to} \quad & \mathbb{E}_{x \sim \mathcal{D}}\Big[\frac{|\mathcal{R}_k(f(x)) \,\cap\, \mathcal{R}_k(G_\theta(f(x)))|}{k}\Big] \ge 1 - \delta_{\text{ret}}.
\end{aligned}
\end{equation}

\section{EntroGuard}

In this section, we provide our privacy-preserving transmission approach, i.e., EntroGuard, to protect embeddings' privacy while maintaining retrieval capabilities without the cloud's additional cooperation. In \cref{sec:Methodology1}, we overview our proposed embedding protection method. Next, we introduce how we generate robust entropy-based perturbation to protect embeddings from being recovered in \cref{sec:Methodology2}. Finally, we introduce the method we implemented to further constrain the intensity of the perturbation to increase the privacy-preserving capability while guaranteeing the retrieval capabilities in \cref{sec:Methodology3}.

\subsection{Overview} \label{sec:Methodology1}

Our main idea is to make the embedding in the recovery process away from the original sensitive text and steer it toward meaningless text. An intuitive idea is to use adversarial sample methods, e.g. FGSM~\cite{goodfellow2014explaining}, PGD~\cite{madry2017towards}, against the recovery process. However, we find that it has poor generality to resist unknown generative models. To this end, we choose to interfere with the essential components of the transformer architecture to achieve robust privacy-preserving embedding generation. Meanwhile, we also impose a strict constraint on the degree of perturbation to maintain the retrieval performance, as the perturbation intensity is the main factor that determines whether it can be retrieved correctly.

The whole process of EntroGuard is illustrated in \cref{fig:pipe}. When the user inputs the original texts into the end devices, they are first converted into raw embeddings by the original embedding model. These embeddings then go through EntroGuard that contains two components, i.e., Entropy-based Perturbation Generation and Bound-aware Perturbation Adaptation, resulting in a protected embedding.

Specifically, the first component is used to perturb raw embeddings to disrupt the recovery process in recovery models. In particular, we maximize the information entropy of intermediate results in transformer blocks, making the embedding generate meaningless words as early as possible during the recovery process. And we also increase cross-entropy to make the recovered text deviate further from the original sensitive text. The second component is used to constrain the strength of the perturbation with the strategy of ``reducing where redundant and increasing where sparse". By reducing redundant perturbation, it preserves essential semantic information in embeddings to guarantee retrieval performance. Meanwhile, by increasing perturbation that is sparse, the protected embeddings will gain a certain degree of randomness to enhance their resilience against EIAs.

\begin{figure}[!t]
    \centering
    \includegraphics[width=0.8\columnwidth]{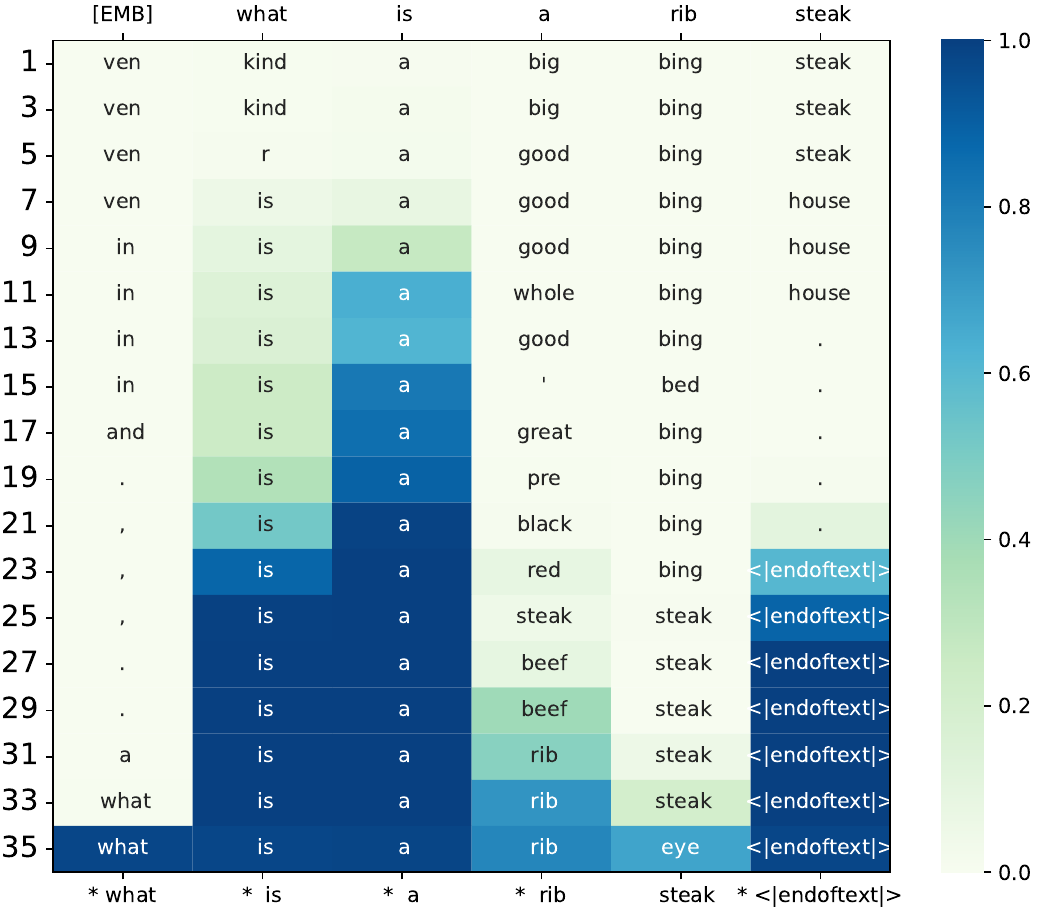}
    \caption{The internal recovery process of EIA, the top is the input, the bottom is the output, and vertical axis represents the Transformer block. The darker the color means the higher the confidence of the generated results, and the words marked with * indicate that the prediction is correct.}
    \label{fig:LL_recover}
\end{figure}

\subsection{Entropy-based Perturbation Generation} \label{sec:Methodology2}

The key to the success of EIAs lies in the powerful generative capabilities of the models they employ. To better prevent EIAs, it is crucial to understand how generative language models recover the original sentence from an embedding. To this end, we use an interpretability tool called Logits Lens~\cite{ll} that can display the output of the intermediate process in the generative model. As illustrated in \cref{fig:LL_recover}, recovery is not always a straightforward process where the correct token is predicted only at the end. Instead, we find that some tokens may be recovered correctly during the middle or even earlier generation stages, which is similar to the normal generative process itself.

This observation led us to the idea of interfering with the text recovery process from the outset to guide the recovery model producing non-sensical words. However, since the generative models deployed by real-world adversaries may vary (e.g., GPT~\cite{ouyang2022training}, Llama~\cite{touvron2023llama}), a defense designed against a specific attack model may fail to withstand attacks from different models. Considering that the mainstream generative language models are always built on the transformer architecture, we perturb embeddings against the fundamental components shared across generative language models, i.e., the transformer block, to make perturbation adaptable to a wide range of generative attack models. Eventually, to achieve privacy protection, we perturb embeddings by increasing the information entropy of intermediate results in transformer blocks to steer them toward meaningless content and increasing the cross-entropy of the final recovered text to keep them away from the original sensitive text.

To implement this, we aim to optimize a perturbation generator $G_\theta$ to perturb embeddings for privacy preservation. First of all, we establish a surrogate model $S$ to approximate the attacker's generation process, which can be any of the mainstream generative language models. Such a surrogate model $S$ is trained with public dataset $\mathcal{X} = \left\{x_{1}, \ldots, x_{n}\right\}$ via \cref{eq:learning attack Loss}, aiming to recover the texts from embeddings, i.e., $S(f(x)) \approx f^{-1}(f(x))=x$.

\begin{figure}[!t]
    \centering
    \includegraphics[width=0.8\columnwidth]{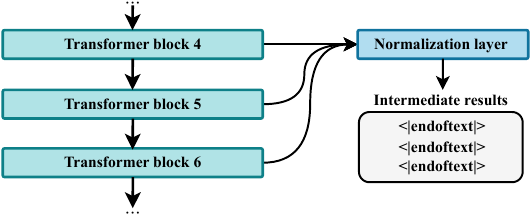}
    \caption{The process of generating intermediate results of each layer of transformer blocks, where the intermediate results converge towards meaningless words via Entropy-based Perturbation Generation.}
    \label{fig:pipe_entropy}
\end{figure}

As illustrated in \cref{fig:pipe_entropy}, in order to supervise the transformer blocks~\cite{ll} at different layers in the recovery process, we leverage the activation layer of the final layer in generative language models to extract the intermediate results, i.e., word distribution $Distr(\cdot)$, of transformer blocks in different layer $k$ throughout the generative process:
\begin{equation}
\begin{split}
\operatorname{Distr}(k)=\operatorname{Softmax}\left(LN_{s}(b_k)  W_{dec}\right) \in \mathbb{R}^{|\mathcal{V}|},
\end{split}
\label{eq:ll_layer}
\end{equation}
\noindent where $LN_{s}(\cdot)$ denotes the final normalization layer of the surrogate model $S$ before the decoding matrix $W_{dec}$, $b_{k}$ represents the output vectors of the $k^{th}$ transformer blocks, and $|\mathcal{V}|$ denotes the size of the vocabulary $\mathcal{V}$.

After obtaining the intermediate results, we increase their information entropy to steer them toward meaningless texts. Specifically, we utilize the average information entropy of the intermediate results as the loss function:
\begin{equation}
\begin{split}
    L_{Entropy} = \frac{1}{L} \sum_{k=1}^{K} \sum_{j=1}^{L} \left( - \sum_{i=1}^{|\mathcal{V}|} p_{k,j}(w_{i}) \log p_{k,j}(w_{i}) \right),
\end{split}
\label{eq:entropy}
\end{equation}

\noindent where $L$ represents the sequence length, $K$ denotes the total number of transformer blocks to be supervised, and $p_{k,j}(w_{i}) \in \operatorname{Distr}(k)$ is the probability of occurrence for the $i$-th word $w_i$ in the vocabulary $\mathcal{V}$ predicted at the $j$-th position within the $k$-th transformer block. This objective calculates the entropy at each token position, averages it across the sequence length $L$ via a mean operation, and subsequently aggregates the results over the $K$ selected blocks. Under this loss function, the supervision from each intermediate result has a cumulative effect on the shallower transformer blocks, resulting in more substantial gradient constraints at shallow blocks, which renders the recovery process ineffective at an early stage. Additionally, we also increase the cross-entropy $L_{CE}$ (shown in \cref{eq:learning attack Loss}) to keep the generated sentence distinct from the original text.

On the other hand, to satisfy the utility goal defined by the theoretical Top-$K$ retrieval constraint $\delta_{ret}$, we must ensure the protected embedding yields retrieval results similar to the original. While directly optimizing for ranking operations like retrieval recall is impractical during training due to their non-differentiable nature and computational complexity, an effective surrogate approach is to minimize the perturbation applied to the embedding itself. Therefore, we constrain the perturbation using a similarity loss $L_{Similarity}$, based on the cosine distance between two embeddings $\mathbf{e_1}$ and $\mathbf{e_2}$:
\begin{equation}
\begin{split}
L_{Similarity} = 1 - \cos(\mathbf{e_1}, \mathbf{e_2}) = 1 -\frac{\mathbf{e_1} \cdot \mathbf{e_2}}{\|\mathbf{e_1}\| \|\mathbf{e_2}\|},
\end{split}
\label{eq:cos}
\end{equation}
where $\mathbf{e_1}$ and $\mathbf{e_2}$ correspond to the original embedding $f(x)$ and the perturbed embedding $G_\theta(f(x))$, respectively.

Eventually, as shown in \cref{fig:intro_pipe}, the optimization objectives for perturbation generator $G_\theta$ to generate entropy-based perturbation on embeddings can be represented as:
\begin{equation}
\begin{split}
\min_{G_\theta} \ Loss = \alpha \cdot L_{Similarly} - \beta \cdot L_{Entropy} - \gamma \cdot L_{CE},
\end{split}
\label{eq:loss_all}
\end{equation}

\noindent where $\alpha$, $\beta$, $\gamma$ are hyperparameters that control the ratios of each loss. Note that while the utility term ($L_{Similarity}$) and the privacy terms ($L_{Entropy}$, $L_{CE}$) introduce opposing gradients, \cref{eq:loss_all} formulates a standard multi-objective optimization problem rather than a self-contradictory one. Guided by the dynamically adjusted hyperparameters, these competing objectives are jointly optimized during training to protect privacy while maintaining retrieval utility.

\subsection{Bound-aware Perturbation Adaptation} \label{sec:Methodology3}

Since the entropy-based perturbation, which achieves privacy protection, may affect retrieval performance, we need to further adapt the perturbation to meet the retrieval requirement simultaneously. As mentioned by Morris et al.~\cite{morris2023text}, there is an existing perturbation bound that both maintains retrieval performance and defends optimization-based EIAs. Although the method they proposed cannot defend against learning-based attacks, as shown in \cref{table:noise_influence}, the retrieval bound $\epsilon$ they measured can still serve as a reference for us. Hence, we propose an algorithm called Bound-aware Perturbation Adaptation to keep the perturbations within the bound, with a strategy of \textit{reducing where redundant and increasing where sparse}. In particular, we reduce perturbations that exceed the bound to ensure retrieval performance while increasing perturbations that do not reach the bound to achieve better protection.

\begin{algorithm}
\caption{Bound-aware Perturbation Adaptation} \label{algorithm:adpat} 
\KwIn{$\mathbf{e^{0}}$: Raw embedding, $\mathbf{e'}$: Perturbed embedding, $\text{dim}(\cdot)$: Total dimension, $\epsilon$: Total perturbation bound, $\lambda$: Scaling factor ($\lambda<1$).}
\KwOut{$\mathbf{e''}$: Adapted perturbed embedding.}

\SetKwData{RD}{random index set}
\SetKwData{RN}{random noise}
\SetKwData{N}{noise}

Compute similarity $L_{Similarity}$: $\rho \gets 1-\cos(\mathbf{e^0}, \mathbf{e'})$;

\If{Perturbation intensity below bound: $\rho < \epsilon$}{
    Uniformly sample a \RD $\{d_1, \ldots, d_k\}$ of $k$ dimensions from $\mathbf{e'}$, where $k = \lfloor \text{dim}(\mathbf{e'})/4 \rfloor$; \label{line1}
    
    Generate \RN $n \in \{n^{+} \sim \mathcal{N}(1, 1)$, $n^{-} \sim \mathcal{N}(-1, 1)\}$ \label{line2} in \RD, where $\mathcal{N}$ denotes the normal distribution;
    
    Initialize adapted embedding: $e'' \gets e'+n$;
    
    \While{$(\rho \gets 1-\cos(e^0,e'')) > \epsilon$}{ \label{line3}
        Decrease the intensity of \N $n$ by $n \gets n \times \lambda$;

        Update embedding: $e'' \gets e'+n$;
    } \label{line4}
}
\Else{

    Initialize adapted embedding: $e'' \gets e'$;

    Initialize noise intensity: $I_{noise} \gets e' - e^0$;

    \While{$(\rho \gets 1-\cos(e^0,e'')) > \epsilon$}{ \label{line5}
      Scale perturbation: $I_{noise} \gets I_{noise} \times \lambda$;
      
      Update embedding: $e'' \gets e^0 + I_{noise}$;
      
    
      } \label{line6}
}
\textbf{Return} Adapted perturbed embedding $e''$\;
\end{algorithm}

As illustrated in \autoref{algorithm:adpat}, if the perturbed embedding $e'$ given by Entropy-based Perturbation Generation is below the predefined bound, we will randomly select a subset of dimensions in the embedding space (\autoref{line1}) and inject a random Gaussian noise (\autoref{line2}) into these dimensions. The initial values of the Gaussian noise are relatively large compared to the embedding and will gradually adjust until the overall perturbation remains within the predefined bound (\autoref{line3} to \autoref{line4}). If the perturbation exceeds the bound, we iteratively scale it down in equal proportion using an exponential decay (\autoref{line5} to \autoref{line6}) until it falls within the acceptable bound. After the above adaptation, all perturbations will remain close to and within the bound, ensuring that the protected embedding retains its retrieval performance without additional cooperation in the cloud.

\section{Experiments}
To evaluate the effectiveness of EntroGuard, we conduct extensive evaluations of retrieval performance and privacy-preserving capability across different embedding models with different datasets. In this section, we will first introduce the setup of our experiments in \cref{sec:Experiments1}, and then we demonstrate the superiority of EntroGuard by answering the following research questions:

\begin{itemize}

\item \textbf{RQ1}: Does EntroGuard influence retrieval performance in the cloud? (\cref{sec:Experiments3})

\item \textbf{RQ2}: Can EntroGuard effectively defend against embedding inversion attacks? (\cref{sec:Experiments2} and \cref{sec:Experiments5})

\item \textbf{RQ3}: Can EntroGuard efficiently process embeddings in the real-world end devices with low costs? (\cref{sec:Experiments4})

\end{itemize}

Specifically, as privacy protection will be meaningless without effective retrieval, we first assess the retrieval performance of our method. Then, we evaluate its privacy protection capabilities and robustness under the premise of maintaining retrieval. Finally, we validate the method's efficiency and deployment feasibility on end devices.

\subsection{Experimental Setup} \label{sec:Experiments1}

\paragraph{Datasets} In our experiments, we evaluate privacy-preserving capability using PersonaChat~\cite{zhang2018personalizing} and MS MARCO~\cite{nguyen2016ms}, partitioned into a 19:1 train-test split to train and test all methods. Moreover, we use ArguAna~\cite{wachsmuth2018retrieval} and FEVER~\cite{thorne2018fever} from the BEIR benchmark~\cite{thakur2021beir} to assess retrieval performance.

\begin{itemize}

\item \textbf{PersonaChat}: a chit-chat dataset consisting of 162,064 utterances, where randomly paired workers are assigned specific personas and engage in conversations to get to know each other.

\item \textbf{MS MARCO}: a question-answering dataset consisting of real Bing queries paired with 8.8 million human-generated answers.

\item \textbf{ArguAna}: a dataset consisting of 8,674 passages, focused on retrieving the most relevant counterarguments for a given argument.

\item \textbf{FEVER}: a fact extraction and verification dataset consisting of 5.42 million passages, designed to support automatic fact-checking systems.

\end{itemize}

\paragraph{The implementation of EntroGuard} In experiments, the generator $G_{\theta}$ is implemented as a custom 1D DenseNet~\cite{huang2017densely} that maps the input embedding $e^0$ to a protected embedding $e'$ of the same dimension ($\mathbb{R}^d \rightarrow \mathbb{R}^d$). The network begins with an initial linear layer that projects the input to a hidden dimension of 256. This is followed by 3 Dense Blocks, each comprising 2 linear layers with a growth rate of 256. Dense connections concatenate the features across blocks to form a 1792-dimensional aggregated representation. Finally, a Dropout layer ($p=0.1$) and a linear projection map the representation back to the original embedding dimension $d$. Because the latent semantic spaces are entirely distinct across different embedding architectures, a separate $G_{\theta}$ is instantiated and trained for each specific host model $f(\cdot)$ (e.g., Sentence-T5, SimCSE-BERT). During training, the host embedding model $f(\cdot)$ remains frozen to preserve its original retrieval utility. However, it remains part of the differentiable computational graph, allowing gradients to backpropagate through $f(\cdot)$ to optimize the parameters of $G_{\theta}$. To train the perturbation generator for 3 epochs, we adopt the Adam optimizer~\cite{kingma2014adam} configured with an initial learning rate of $3\times10^{-5}$ and a weight decay of 0.01 (excluding parameters related to ``bias'', ``ln'', and ``LayerNorm.weight'' terms). To balance competing objectives and ensure optimization stability, the weighting coefficients undergo dynamic adjustment via a constant warmup followed by a cosine annealing strategy. For the initial $100$ warmup steps, the coefficients are fixed at $\alpha = \alpha_{\max}$ and $\beta = \beta_{\min}$. For subsequent iterations $m$ (out of $M$ total iterations), they are updated as $\alpha^{(m)} = \alpha_{\min} + \frac{1}{2}(\frac{\alpha_{\max}}{2} - \alpha_{\min})(1 + \cos(\pi \frac{m}{M}))$ and $\beta^{(m)} = \beta_{\max} - \frac{1}{2}(\beta_{\max} - \beta_{\min})(1 + \cos(\pi \frac{m}{M}))$, where we set $\alpha_{\max}=1200$, $\alpha_{\min}=300$, $\beta_{\max}=2$, and $\beta_{\min}=1$ (with $\gamma$ fixed at $1$). Additionally, if not specified in the experiments, $\lambda$ is set to 0.95, and $\epsilon$ is set to 0.036, which corresponds to the intensity of 0.01x Gaussian noise with a mean of 0 and a variance of 1.

\paragraph{The implementation of surrogate model} The surrogate model is instantiated using pre-trained causal language models (e.g., DialoGPT-large) accompanied by an initial linear projection layer to align the dimensions of the host embeddings (e.g., mapping 768 or 1024 dimensions) with the generative model's hidden states (e.g., 1280 dimensions). Similarly, the model is trained for 5 epochs utilizing the AdamW optimizer with a learning rate of $3 \times 10^{-5}$, an epsilon of $1 \times 10^{-6}$, and a weight decay of $0.01$ (excluding bias and LayerNorm parameters), employing a linear learning rate scheduler with 100 warmup steps.

\paragraph{The implementation of Attack methods} In our experiments, we adopt the representative work of learning-based EIAs, i.e., GEIA~\cite{li2023sentence}, as the primary attack method due to its strong privacy attack capability. To rigorously evaluate EntroGuard under a worst-case scenario, we assume the adversary has access to a public corpus $\mathcal{X}$ that follows the same distribution as the user's private data $\mathcal{D}$. Meanwhile, we independently train a separate attack model for each host embedding model (e.g., Sentence-T5, MPNet) and target dataset (e.g., PersonaChat, MS MARCO) to ensure the evaluation reflects the strongest possible threat. For the GEIA decoding configuration, we utilize a beam width of 5. At each step $t$, every active beam is expanded with its top-5 next-token candidates to find the sequence with the highest cumulative log-probability. The maximum decoding length is capped at $L_{\max}=50$ tokens, and the autoregressive decoding process terminates only when this maximum generation step is reached. Crucially, before calculating any evaluation metrics (e.g., ROUGE, BLEU), we strip all generated \texttt{<|endoftext|>} tokens from the outputs to ensure the leakage metrics accurately reflect the recovered semantic content rather than decoding artifacts. Moreover, we use the representative work of optimization-based attacks, i.e., Vec2Text~\cite{morris2023text} for evaluation. Note that we utilize the official implementations for both methods, including their default training configurations and hyperparameters, such as training for 10 epochs with a batch size of 16.

\paragraph{Embedding models \& Recovery models} For embedding models on the end device, we employ widely used models, including Sentence-T5~\cite{ni2022sentence}, SimCSE-BERT~\cite{gao2021simcse}, RoBERTa~\cite{liu2019roberta}, and MPNet~\cite{song2020mpnet} for learning-based EIAs, as well as GTR-T5~\cite{ni2022large} for optimization-based EIAs, all utilizing their officially trained versions. For adversaries' recovery models, we utilize GPT-2~\cite{radford2019language}, DialoGPT~\cite{zhang2019dialogpt}, and Llama3~\cite{dubey2024llama} in learning-based EIAs, which are all trained on datasets from the same domain as the victim data to ensure optimal performance.

\paragraph{Baseline Defense Methods} We compare our method with three related protection methods with different mechanisms, i.e., Differential Privacy~\cite{li2021deepobfuscator}, PGD~\cite{madry2017towards}, Textobfuscator~\cite{zhou2023textobfuscator}. Specifically, Differential Privacy adds random Gaussian noise to embeddings to protect embedding privacy. Following the setting in \cite{morris2023text}, we use a privacy budget of 0.01. PGD utilizes projected gradient descent to add adversarial noise to the embedding and protect privacy by disturbing specific attack models. In this method, we also use a noise strength consistent with DP, i.e., 0.01. Textobfuscator ensures inference privacy by introducing perturbations to the clustered embeddings. To better adapt to our scenario, we transformed its task objective from the original classification tasks to retrieval tasks.

\begin{table*}[!t]
  \centering
  \caption{The performance of retrieval capability in cloud database in terms of NDCG, MAP, and Precision with different embedding models and methods on the end side. The higher the values of NDCG, MAP, and Precision, the better the retrieval performance.}
  \resizebox{1.95\columnwidth}{!}{
    \begin{tabular}{llccc|ccc|ccc}
    \toprule
    \multirow{2}[2]{*}{\begin{tabular}[c]{@{}c@{}}{Embedding}\\{model}\end{tabular}} & \multicolumn{1}{c}{\multirow{2}[2]{*}{Method}} & \multicolumn{3}{c}{ArguAna} & \multicolumn{3}{c}{FEVER} & \multicolumn{3}{c}{MS MARCO} \\
\cmidrule{3-11}          &       & NDCG↑ & MAP↑  & Precision↑ & NDCG↑ & MAP↑  & Precision↑ & NDCG↑ & MAP↑  & Precision↑ \\
    \midrule
    \multirow{5}[2]{*}{Sentence-t5} 
          & Original & 0.3364  & 0.2853  & 0.0983  & 0.3220  & 0.2832  & 0.0915  & 0.5084  & 0.0496  & 0.5954  \\
          & DP & 0.3362  & 0.2849  & 0.0984  & 0.3127  & 0.2746  & 0.0891  & 0.5095  & 0.0529  & 0.5861  \\
          & PGD & 0.3284  & 0.2802  & 0.0949  & 0.2983  & 0.2616  & 0.0853  & 0.4914  & 0.0486  & 0.5767  \\
          & TextObfuscator & 0.3385  & 0.2865  & 0.0992  & 0.3453  & 0.3049  & 0.0972  & 0.5016  & 0.0481  & 0.5907  \\
          & \cellcolor[rgb]{ .851,  .851,  .851}Ours & \cellcolor[rgb]{ .851,  .851,  .851}0.3464  & \cellcolor[rgb]{ .851,  .851,  .851}0.2937  & \cellcolor[rgb]{ .851,  .851,  .851}0.1011  & \cellcolor[rgb]{ .851,  .851,  .851}0.3237  & \cellcolor[rgb]{ .851,  .851,  .851}0.2867  & \cellcolor[rgb]{ .851,  .851,  .851}0.0908  & \cellcolor[rgb]{ .851,  .851,  .851}0.4933  & \cellcolor[rgb]{ .851,  .851,  .851}0.0478  & \cellcolor[rgb]{ .851,  .851,  .851}0.5954  \\
    \midrule
    \multirow{5}[2]{*}{Simcse-bert} 
          & Original & 0.3520  & 0.3010  & 0.1013  & 0.1621  & 0.1394  & 0.0478  & 0.2848  & 0.0319  & 0.4000  \\
          & DP & 0.3511  & 0.3000  & 0.1011  & 0.1623  & 0.1394  & 0.0479  & 0.2832  & 0.0320  & 0.4000  \\
          & PGD & 0.3616  & 0.3123  & 0.1020  & 0.1291  & 0.1101  & 0.0388  & 0.2887  & 0.0359  & 0.3954  \\
          & TextObfuscator & 0.3427  & 0.2936  & 0.0982  & 0.1717  & 0.1489  & 0.0498  & 0.2736  & 0.0295  & 0.3861  \\
          & \cellcolor[rgb]{ .851,  .851,  .851}Ours & \cellcolor[rgb]{ .851,  .851,  .851}0.3484  & \cellcolor[rgb]{ .851,  .851,  .851}0.2988  & \cellcolor[rgb]{ .851,  .851,  .851}0.0997  & \cellcolor[rgb]{ .851,  .851,  .851}0.1690  & \cellcolor[rgb]{ .851,  .851,  .851}0.1455  & \cellcolor[rgb]{ .851,  .851,  .851}0.0497  & \cellcolor[rgb]{ .851,  .851,  .851}0.2651  & \cellcolor[rgb]{ .851,  .851,  .851}0.0304  & \cellcolor[rgb]{ .851,  .851,  .851}0.3674  \\
    \midrule
    \multirow{5}[2]{*}{MPNet} 
          & Original & 0.4534  & 0.3948  & 0.1260  & 0.5638  & 0.5080  & 0.1515  & 0.6928  & 0.0904  & 0.8140  \\
          & DP & 0.4532  & 0.3944  & 0.1262  & 0.5612  & 0.5048  & 0.1514  & 0.6834  & 0.0893  & 0.8000  \\
          & PGD & 0.4543  & 0.3964  & 0.1259  & 0.5580  & 0.5029  & 0.1498  & 0.6991  & 0.0892  & 0.8140  \\
          & TextObfuscator & 0.4607  & 0.4018  & 0.1277  & 0.5780  & 0.5233  & 0.1537  & 0.7074  & 0.0915  & 0.8140  \\
          & \cellcolor[rgb]{ .851,  .851,  .851}Ours & \cellcolor[rgb]{ .851,  .851,  .851}0.4517  & \cellcolor[rgb]{ .851,  .851,  .851}0.3946  & \cellcolor[rgb]{ .851,  .851,  .851}0.1248  & \cellcolor[rgb]{ .851,  .851,  .851}0.5592  & \cellcolor[rgb]{ .851,  .851,  .851}0.5038  & \cellcolor[rgb]{ .851,  .851,  .851}0.1502  & \cellcolor[rgb]{ .851,  .851,  .851}0.6953  & \cellcolor[rgb]{ .851,  .851,  .851}0.0901  & \cellcolor[rgb]{ .851,  .851,  .851}0.8186  \\
    \midrule
    \multirow{5}[2]{*}{Roberta} 
          & Original & 0.3852  & 0.3293  & 0.1108  & 0.5011  & 0.4521  & 0.1344  & 0.6706  & 0.0729  & 0.7861  \\
          & DP & 0.3871  & 0.3300  & 0.1120  & 0.4976  & 0.4483  & 0.1339  & 0.6717  & 0.0733  & 0.7861  \\
          & PGD & 0.3799  & 0.3225  & 0.1108  & 0.4932  & 0.4449  & 0.1324  & 0.6715  & 0.0732  & 0.7861  \\
          & TextObfuscator & 0.3999  & 0.3425  & 0.1147  & 0.5262  & 0.4766  & 0.1402  & 0.6726  & 0.0736  & 0.7861  \\
          & \cellcolor[rgb]{ .851,  .851,  .851}Ours & \cellcolor[rgb]{ .851,  .851,  .851}0.3837  & \cellcolor[rgb]{ .851,  .851,  .851}0.3271  & \cellcolor[rgb]{ .851,  .851,  .851}0.1110  & \cellcolor[rgb]{ .851,  .851,  .851}0.4878  & \cellcolor[rgb]{ .851,  .851,  .851}0.4387  & \cellcolor[rgb]{ .851,  .851,  .851}0.1317  & \cellcolor[rgb]{ .851,  .851,  .851}0.6736  & \cellcolor[rgb]{ .851,  .851,  .851}0.0738  & \cellcolor[rgb]{ .851,  .851,  .851}0.7814  \\
    \bottomrule
    \end{tabular}}%
  \label{tab:retrieval}%
\end{table*}

\paragraph{Evaluation metrics} For evaluating the privacy-preserving capability of the embedding, we utilize metrics that compare text similarity, i.e., ROUGE~\cite{lin2004rouge}, BLEU~\cite{papineni2002bleu}, EMR (Exact Match)~\cite{morris2023text}. Higher text similarity indicates more privacy leakage. Specifically, ROUGE is a metric that evaluates the overlap of unigrams (individual words) between the system-generated summary and the reference summary. BLEU is an algorithm designed to assess the quality of text that has been machine-translated from one natural language to another, comparing the output to reference translations. EMR (Exact Match) quantifies the percentage of generated outputs that exactly match the ground truth. In the following experiments, we specifically utilize ROUGE-2 and BLEU-2.

However, semantic leakage in text also constitutes a form of privacy leakage. Since the above metrics mainly assess privacy leakage based on text similarity (e.g., n-gram matching), they may not effectively capture deeper semantic similarities in certain cases. For example, two sentences may be nearly the same in vocabulary but express different meanings, in which case ROUGE and BLEU will give a high similarity score. In order to measure whether privacy has been compromised from the semantic similarity, we propose a new metric, BiNLI. Inspired by Kuhn et al.~\cite{kuhn2023semantic}, we use the Natural Language Inference (NLI) Model to determine the entailment relationship between two statements to reflect their semantic similarity. Specifically, BiNLI can be calculated as:
\begin{equation}
\begin{split}
\operatorname{BiNLI} = \frac{\sum_{i=1}^{N}[\operatorname{entail}(a_{i},b_{i}) + \operatorname{entail}(b_{i},a_{i})]}{2N}
\end{split}
\label{eq:binli}
\end{equation}

\noindent where $N = |\mathcal{A}| = |\mathcal{B}|$ represents the total number of sample pairs in datasets $\mathcal{A}=\{a_1,a_2,\dots,a_N\}$ and $\mathcal{B}=\{b_1,b_2,\dots,b_N\}$. The sentences $a_i$ and $b_i$ are considered semantically related when $a_i$ entails $b_i$ or $b_i$ entails $a_i$, which we quantify using a hard-labeling function $\operatorname{entail}(\cdot, \cdot)$ that takes the argmax over the output logits to output $1$ for the entailment class and $0$ otherwise. In our experiments, we implement this scoring function using the pre-trained microsoft/deberta-large-mnli checkpoint to compute the proposed BiNLI for measuring the privacy leakage of texts, with higher scores indicating greater leakage.

Moreover, to evaluate retrieval performance, we use NDCG to assess how well the actual retrieval ranking aligns with the ideal order, MAP to assess average precision by considering both the number of relevant knowledge and their position in the retrieval, and Precision to measure the proportion of relevant results in retrieved knowledge. Specifically, we utilize NDCG@5, MAP@5, and Precision@5 in our experiments.

The datasets utilized in this study are publicly accessible. The source code for reproducing the EntroGuard framework is available at \url{https://github.com/shuaifanjin/EntroGuard}. All experiments are conducted on NVIDIA RTX 5880 GPUs with a random seed of 1024.

\begin{table*}[!ht]
  \centering
  \caption{The examples of privacy-preserving capabilities against learning-based EIAs. The lower the values of BLEU, BiNLI, the better the privacy-preserving performance.}

  \resizebox{1.95\columnwidth}{!}{

    \begin{tabular}{llcc}
    \toprule
    User's Text & what can you eat when you have an attack of diarrhea & BLEU  & BiNLI \\
    \midrule
    Original & what can you eat when you have diarrhea\texttt{\texttt{<|endoftext|>}} & 0.6363  & 1.00  \\
    DP    & what can you eat when you have severe diarrhea\texttt{<|endoftext|>} & 0.6538  & 1.00  \\
    PGD   & what can diarrhea meal be\texttt{<|endoftext|>} & 0.1167  & 0.50  \\
    TextObfuscator & what can you eat when you have diarrhea\texttt{<|endoftext|>} & 0.6363  & 1.00  \\
    \cellcolor[rgb]{ .851,  .851,  .851}Ours  & \cellcolor[rgb]{ .851,  .851,  .851}\texttt{<|endoftext|>}\texttt{<|endoftext|>}\texttt{<|endoftext|>}\texttt{<|endoftext|>}\texttt{<|endoftext|>} & \cellcolor[rgb]{ .851,  .851,  .851}0.0000  & \cellcolor[rgb]{ .851,  .851,  .851}0.00  \\
    \midrule
    \midrule
    User's Text & i thought you said you would take up gaming , you know i am a gamer & BLEU  & BiNLI \\
    \midrule
    Original & oh i thought you said you were into gaming. i will become a gamer\texttt{<|endoftext|>} & 0.4064  & 0.50  \\
    DP    & i thought you said you were into gaming . i will play .\texttt{<|endoftext|>} & 0.3364  & 0.50  \\
    PGD   & oh okay . i thought you said you play guitar . i am into baking .\texttt{<|endoftext|>} & 0.3819  & 0.00  \\
    TextObfuscator & oh i thought you said you were a gamer . i am into gaming .\texttt{<|endoftext|>} & 0.5000  & 0.50  \\
    \cellcolor[rgb]{ .851,  .851,  .851}Ours  & \cellcolor[rgb]{ .851,  .851,  .851}\texttt{<|endoftext|>}\texttt{<|endoftext|>}\texttt{<|endoftext|>}\texttt{<|endoftext|>}\texttt{<|endoftext|>} & \cellcolor[rgb]{ .851,  .851,  .851}0.0000  & \cellcolor[rgb]{ .851,  .851,  .851}0.00  \\

    \bottomrule
    \end{tabular}}
  \label{tab:Privacy-preserving results}%
\end{table*}%

\subsection{Retrieval performance} \label{sec:Experiments3}
In this section, we compare the retrieval performance of our proposed EntroGuard with original embedding models of different architectures.

\textbf{EntroGuard can maintain high retrieval performance.} To protect end-side privacy from the honest but curious server that can invade users' privacy, we retrieve the original knowledge embeddings with protected embeddings from the end side. Specifically, our methods are trained with MS MARCO and evaluated by various datasets, i.e., ArguAna, FEVER, and MS MARCO.

\begin{figure}[!t]
\centering
\includegraphics[width=0.95\columnwidth]{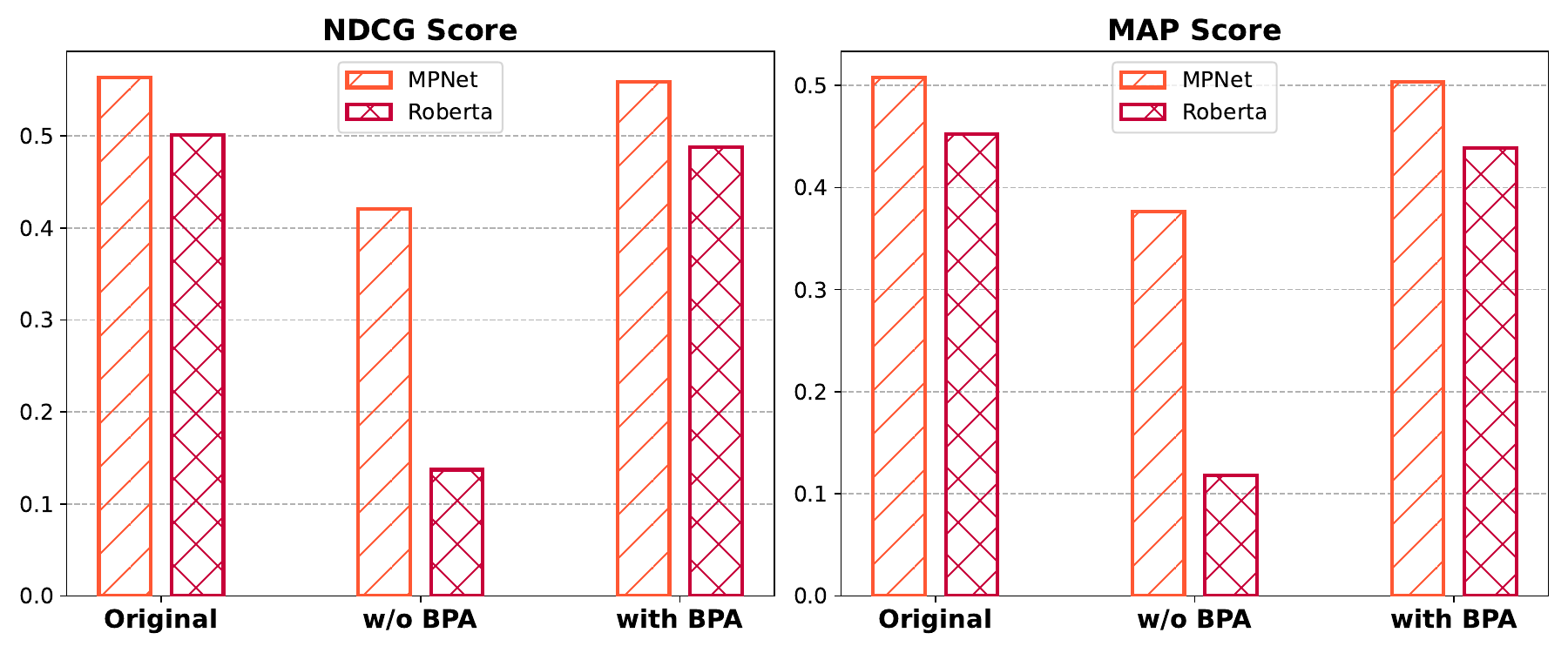}
\caption{The retrieval performance on FEVER dataset.}
\label{fig:comparison_retr}

\end{figure}

Since retrieval performance is primarily related to the similarity between the relevant query and the corpus, i.e., the similarity at the embedding (vector) level, we set the bound $\epsilon$ in our methods to 0.036 to maintain the retrieval performance based on the retrieval bound analysis by Morris et al.~\cite{morris2023text} and Chen et al.~\cite{chen2024text}. \autoref{tab:retrieval} shows the results of retrieval performance. All the metrics of our method remain close to that of the original models (e.g., Sentence-T5: 0.3464 vs. 0.3364, SimCSE-BERT: 0.3484 vs. 0.3520), whose results demonstrate that EntroGuard effectively preserves retrieval performance across different embedding models. The retrieval results of other schemes can also be seen in \autoref{tab:retrieval}, whose results also align with the analysis that the retrieval performance can be maintained by controlling the perturbation intensity. Therefore, the key to evaluation lies in whether these approaches can achieve privacy protection under the condition that retrieval performance is maintained.

Besides, ~\autoref{fig:comparison_retr} shows the role of Bound-aware Perturbation Adaptation (BPA) in maintaining retrieval performance. Without BPA, there is a significant drop in retrieval metrics, and with BPA, the retrieval metrics are almost identical to those of the original embedding model, demonstrating the effectiveness of BPA in maintaining retrieval performance.

\begin{table*}[!ht]
  \centering
  \caption{The performance of privacy-preserving capabilities against learning-based EIAs in terms of ROUGE, BLEU, EMR, and BiNLI with different embedding models. Results are evaluated using the GPT2-Large attacker with Beam Search Decoding. The lower the values of ROUGE, BLEU, EMR, and BiNLI, the better the privacy-preserving performance.}

  \resizebox{1.95\columnwidth}{!}{
    \begin{tabular}{llcccc|cccc}
    \toprule
    \multirow{2}[2]{*}{\begin{tabular}[c]{@{}c@{}}{Embedding}\\{model}\end{tabular}} & \multicolumn{1}{c}{\multirow{2}[2]{*}{Method}} & \multicolumn{4}{c}{PersonaChat} & \multicolumn{4}{c}{MS MARCO} \\
\cmidrule{3-10}          &       & ROUGE↓ & BLEU↓ & EMR↓  & BiNLI↓ & ROUGE↓ & BLEU↓ & EMR↓  & BiNLI↓ \\
    \midrule
    \multirow{5}[2]{*}{Sentence-t5} & Original & 0.4620  & 0.3253  & 0.1841  & 0.5608  & 0.6427  & 0.6907  & 0.3377  & 0.7370  \\
          & DP    & 0.3969  & 0.4309  & 0.1430  & 0.4783  & 0.5439  & 0.6045  & 0.2205  & 0.6450  \\
          & PGD   & 0.0959  & 0.1413  & 0.0002  & 0.1860  & 0.0943  & 0.1525  & 0.0000  & 0.1676  \\
          & TextObfuscator & 0.4599  & 0.4894  & 0.1776  & 0.5630  & 0.6396  & 0.6869  & 0.3261  & 0.7360  \\
          & \cellcolor[rgb]{ .851,  .851,  .851}Ours & \cellcolor[rgb]{ .851,  .851,  .851}\textbf{0.0002 } & \cellcolor[rgb]{ .851,  .851,  .851}\textbf{0.0015 } & \cellcolor[rgb]{ .851,  .851,  .851}\textbf{0.0000 } & \cellcolor[rgb]{ .851,  .851,  .851}\textbf{0.0091 } & \cellcolor[rgb]{ .851,  .851,  .851}\textbf{0.0000 } & \cellcolor[rgb]{ .851,  .851,  .851}\textbf{0.0000 } & \cellcolor[rgb]{ .851,  .851,  .851}\textbf{0.0000 } & \cellcolor[rgb]{ .851,  .851,  .851}\textbf{0.0008 } \\
    \midrule
    \multirow{5}[2]{*}{Simcse-bert} & Original & 0.4695  & 0.5155  & 0.1897  & 0.5619  & 0.6272  & 0.6798  & 0.3197  & 0.6432  \\
          & DP    & 0.4684  & 0.5148  & 0.1890  & 0.5627  & 0.6264  & 0.6796  & 0.3187  & 0.6433  \\
          & PGD   & 0.3082  & 0.3612  & 0.0422  & 0.4205  & 0.3991  & 0.4597  & 0.0411  & 0.4415  \\
          & TextObfuscator & 0.4139  & 0.4678  & 0.1446  & 0.5126  & 0.6080  & 0.6702  & 0.2965  & 0.6373  \\
          & \cellcolor[rgb]{ .851,  .851,  .851}Ours & \cellcolor[rgb]{ .851,  .851,  .851}\textbf{0.0068 } & \cellcolor[rgb]{ .851,  .851,  .851}\textbf{0.0166 } & \cellcolor[rgb]{ .851,  .851,  .851}\textbf{0.0000 } & \cellcolor[rgb]{ .851,  .851,  .851}\textbf{0.0338 } & \cellcolor[rgb]{ .851,  .851,  .851}\textbf{0.0018 } & \cellcolor[rgb]{ .851,  .851,  .851}\textbf{0.0028 } & \cellcolor[rgb]{ .851,  .851,  .851}\textbf{0.0000 } & \cellcolor[rgb]{ .851,  .851,  .851}\textbf{0.0618 } \\
    \midrule
    \multirow{5}[2]{*}{MPNet} & Original & 0.4040  & 0.4429  & 0.1564  & 0.4518  & 0.5642  & 0.6223  & 0.2479  & 0.6776  \\
          & DP    & 0.3812  & 0.4195  & 0.1434  & 0.4243  & 0.5222  & 0.5875  & 0.2040  & 0.6433  \\
          & PGD   & 0.1456  & 0.1350  & 0.0017  & 0.2063  & 0.1101  & 0.1768  & 0.0002  & 0.2389  \\
          & TextObfuscator & 0.3977  & 0.2851  & 0.1508  & 0.4479  & 0.5533  & 0.6176  & 0.2333  & 0.6770  \\
          & \cellcolor[rgb]{ .851,  .851,  .851}Ours & \cellcolor[rgb]{ .851,  .851,  .851}\textbf{0.0131 } & \cellcolor[rgb]{ .851,  .851,  .851}\textbf{0.0017 } & \cellcolor[rgb]{ .851,  .851,  .851}\textbf{0.0012 } & \cellcolor[rgb]{ .851,  .851,  .851}\textbf{0.0191 } & \cellcolor[rgb]{ .851,  .851,  .851}\textbf{0.0314 } & \cellcolor[rgb]{ .851,  .851,  .851}\textbf{0.0432 } & \cellcolor[rgb]{ .851,  .851,  .851}\textbf{0.0024 } & \cellcolor[rgb]{ .851,  .851,  .851}\textbf{0.1386 } \\
    \midrule
    \multirow{5}[2]{*}{Roberta} & Original & 0.3679  & 0.3970  & 0.1360  & 0.4064  & 0.5370  & 0.5978  & 0.2168  & 0.7097  \\
          & DP    & 0.3738  & 0.3407  & 0.1233  & 0.3738  & 0.5577  & 0.4902  & 0.1744  & 0.6676  \\
          & PGD   & 0.0914  & 0.1378  & 0.0004  & 0.1445  & 0.0635  & 0.1128  & 0.0000  & 0.1530  \\
          & TextObfuscator & 0.3586  & 0.3955  & 0.1293  & 0.3988  & 0.5163  & 0.5863  & 0.1915  & 0.7060  \\
          & \cellcolor[rgb]{ .851,  .851,  .851}Ours & \cellcolor[rgb]{ .851,  .851,  .851}\textbf{0.0001 } & \cellcolor[rgb]{ .851,  .851,  .851}\textbf{0.0000 } & \cellcolor[rgb]{ .851,  .851,  .851}\textbf{0.0000 } & \cellcolor[rgb]{ .851,  .851,  .851}\textbf{0.0008 } & \cellcolor[rgb]{ .851,  .851,  .851}\textbf{0.0033 } & \cellcolor[rgb]{ .851,  .851,  .851}\textbf{0.0084 } & \cellcolor[rgb]{ .851,  .851,  .851}\textbf{0.0001 } & \cellcolor[rgb]{ .851,  .851,  .851}\textbf{0.0801 } \\
    \bottomrule
    \end{tabular}}
  \label{tab:Privacy-preserving capabilities}%
\end{table*}%

\subsection{Privacy-preserving capabilities} \label{sec:Experiments2}
In this section, we first show the reliability of the BiNLI metrics in certain situations by analyzing examples. Then, we demonstrate the privacy-preserving capabilities of our method across different embedding models.

\textbf{Metric BiNLI can reflect privacy leakage between sentences from semantic similarity.} The reconstructed sentences in \autoref{tab:Privacy-preserving results} demonstrate two scenarios in which only utilizing text similarity to measure the degree of privacy leakage may fail. The first scenario involves cases where the text has undergone significant changes, but the semantics remain largely intact. For example, in the sentence \textit{``what can diarrhea meal be?"}, the words "diarrhea" and "meal" still convey sensitive information, similar to the original phrase "attack of diarrhea" and "eat". In this case, traditional metrics like BLEU may yield a relatively low score, indicating little privacy disclosure in terms of text similarity. However, a BiNLI score of 0.5 reflects the privacy leakage in this scenario more accurately, indicating that partial privacy has been compromised in terms of semantic similarity. The second scenario, in contrast, occurs when textual changes are minimal, but the semantics shift substantially. For instance, in \textit{``I thought you said you play guitar. i am into baking"}, the keywords ``guitar" and ``baking" differ substantially from the original keywords ``gaming" and ``gamer". Here, traditional metrics might remain high scores, potentially misrepresenting the privacy risk. Conversely, our proposed BiNLI score of 0 effectively reflects the lack of semantic overlap, indicating that privacy has been preserved in terms of semantic similarity. 

In conclusion, by evaluating privacy leakage from the perspective of semantics, our BiNLI metric can serve as a complement to the metrics that only compare similarity from text level, providing a more comprehensive assessment of textual privacy leakage.

\textbf{EntroGuard can efficiently defend against EIAs.} To protect users' privacy, TextObfuscator requires fine-tuning the entire embedding model, whereas DP, PGD, and our proposed EntroGuard can directly perturb the embeddings. Specifically, we evaluate EntroGuard against other privacy protection methods using the MS MARCO and PersonaChat datasets, employing embedding models with different structures, i.e., sentence-t5-large, simcse-bert-large-uncased, all-roberta-large, and all-mpnet-base.

\begin{figure}[!t]
\centering
\includegraphics[width=0.95\columnwidth]{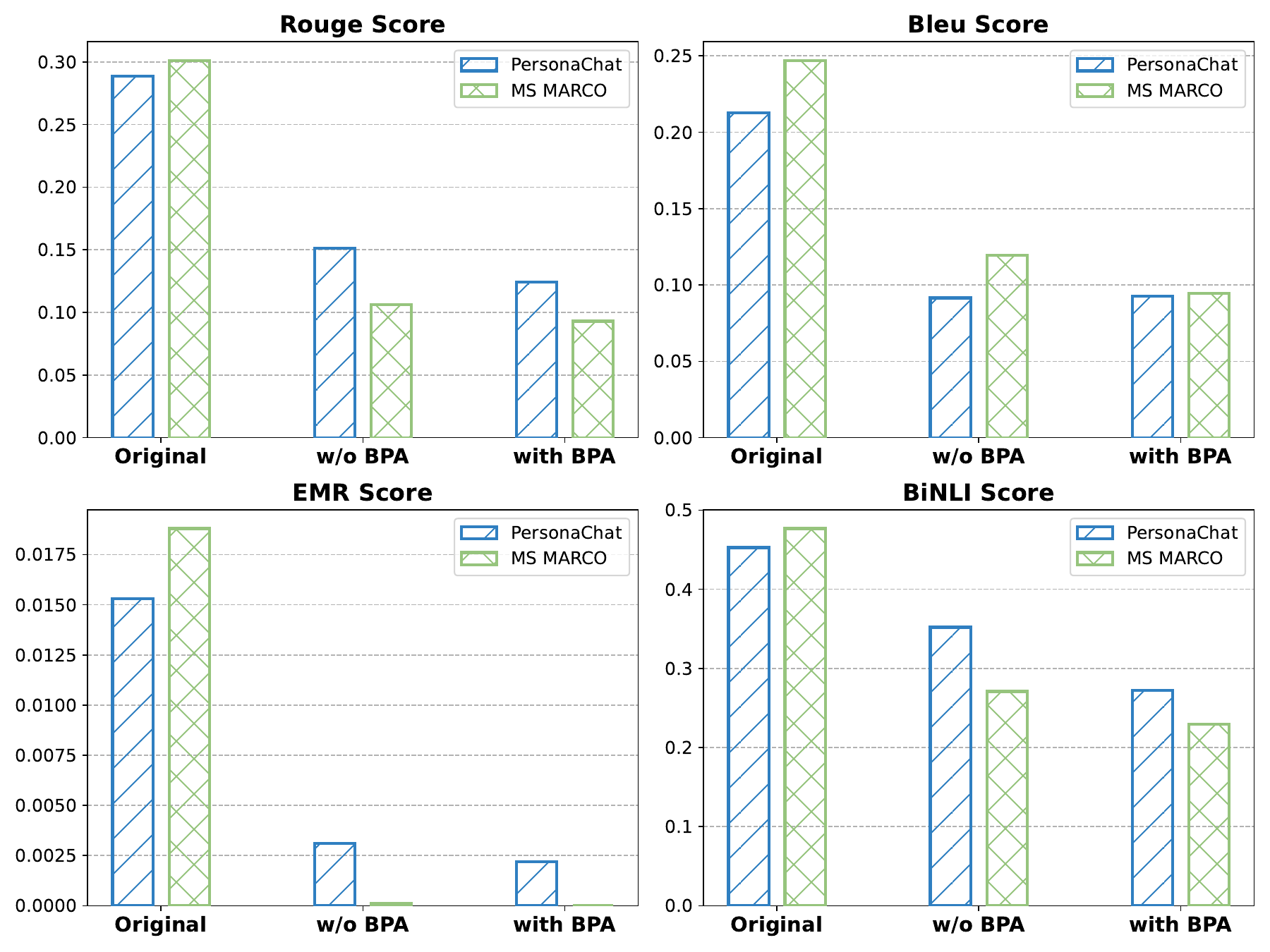}

\caption{The performance of privacy-preserving capabilities against optimization-based EIAs.}
\label{fig:optimize}

\end{figure}

\autoref{tab:Privacy-preserving capabilities} shows the performance of privacy-preserving capabilities against learning-based EIAs. As observed, both the differential privacy-based scheme and TextObfuscator exhibit similar defenses to the original embedding model on text similarity metrics like ROUGE, BLEU, EMR, and the semantic similarity metric, BiNLI. These results demonstrate that these schemes cannot effectively counter learning-based embedding inversion attacks, which still lead to the exposure of user privacy. Although the PGD scheme offers some resistance to embedding inversion attacks, it may still result in partial leakage of user privacy. In particular, in some embedding models, e.g., at simcse-bert, it offers poor privacy protection, with ROUGE only decreasing from 0.4695 to 0.3082 and BiNLI decreasing from 0.5619 to 0.4205. In contrast, the evaluation results for EntroGuard show a substantial reduction in metric values, i.e., ROUGE drops from approximately 0.509 to 0.007, BLEU from 0.534 to 0.009, EMR from 0.224 to 0, and BiNLI from 0.594 to 0.043, which demonstrates that EntroGuard effectively protects user privacy.

\textbf{EntroGuard can resist other types of attacks to some extent.} We also evaluate the effectiveness of EntroGuard against optimization-based EIAs, although these iterative attacks are less practical for real-world exploitation due to their heavy query requirements. Since EntroGuard shifts the target embedding coordinates, it guides the continuous optimization process toward a perturbed anchor rather than the true representation. As demonstrated in \autoref{fig:optimize}, EntroGuard does not suppress optimization-based EIAs as completely as learning-based EIAs. Nevertheless, it still substantially mitigates these attacks. Specifically, on the MS MARCO dataset, with BPA enabled, the BiNLI score decreases from 0.4768 to 0.2292 compared with the baseline, corresponding to a relative reduction of about 51.9\%. For text-level metrics, the ROUGE and BLEU scores fall to 0.0930 and 0.0946, respectively, while the Exact Match Rate reaches 0.0000.

\subsection{Robustness of EntroGuard} \label{sec:Experiments5}
In this section, we evaluate the robustness of EntroGuard. We first verify its defense effectiveness at the logit level and under different decoding strategies. Then, we demonstrate its applicability across heterogeneous end-device embedding models and its generality to different attack models. Finally, we conduct a sensitivity analysis of the scaling factor $\lambda$ and an ablation study on the loss components.

\begin{table}[!t]
\centering
\caption{Logit-level leakage evaluation of sensitive tokens during autoregressive generation.}
\label{tab:logit_leakage}
\resizebox{0.95\columnwidth}{!}{
\begin{tabular}{llc}
\toprule
Metric Category & Metric Name & Score \\
\midrule
\multirow{5}{*}{Top-$k$ Presence} 
& Token Recall@1   & 0.0000 \\
& Token Recall@5   & 0.0007 \\
& Token Recall@10  & 0.0014 \\
& Token Recall@50  & 0.0065 \\
& Token Recall@100 & 0.0146 \\
\midrule
\multirow{1}{*}{Probability Mass} 
& Avg. Probability & 0.000158 \\
\midrule
\multirow{2}{*}{Rank Metrics}     
& Mean Rank        & 12641.71 \\
& Mean Reciprocal Rank & 0.001064 \\
\bottomrule
\end{tabular}}
\end{table}

\begin{table}[!t]
  \centering
  \caption{The applicability of EntroGuard with different size of embedding models.}

  \resizebox{0.95\columnwidth}{!}{
    \begin{tabular}{clcccc}
    \toprule
    Size  & \multicolumn{1}{c}{Method} & ROUGE & BLEU  & EMR   & BiNLI \\
    \midrule
    \multirow{2}[2]{*}{Base} & Original & 0.6417  & 0.6901  & 0.3324  & 0.7287  \\
          & \cellcolor[rgb]{ .851,  .851,  .851}Ours & \cellcolor[rgb]{ .851,  .851,  .851}0.0000  & \cellcolor[rgb]{ .851,  .851,  .851}0.0000  & \cellcolor[rgb]{ .851,  .851,  .851}0.0000  & \cellcolor[rgb]{ .851,  .851,  .851}0.0004  \\
    \midrule
    \multirow{2}[1]{*}{Large} & Original & 0.6427  & 0.6907  & 0.3377  & 0.7370  \\
          & \cellcolor[rgb]{ .851,  .851,  .851}Ours & \cellcolor[rgb]{ .851,  .851,  .851}0.0000  & \cellcolor[rgb]{ .851,  .851,  .851}0.0000  & \cellcolor[rgb]{ .851,  .851,  .851}0.0000  & \cellcolor[rgb]{ .851,  .851,  .851}0.0008  \\
    \midrule
    \multirow{2}[1]{*}{XL} & Original & 0.6186  & 0.6686  & 0.3117  & 0.7280  \\
          & \cellcolor[rgb]{ .851,  .851,  .851}Ours & \cellcolor[rgb]{ .851,  .851,  .851}0.0000  & \cellcolor[rgb]{ .851,  .851,  .851}0.0000  & \cellcolor[rgb]{ .851,  .851,  .851}0.0000  & \cellcolor[rgb]{ .851,  .851,  .851}0.0004  \\
    \midrule
    \multirow{2}[2]{*}{XXL} & Original & 0.6133  & 0.6133  & 0.3039  & 0.7227  \\
          & \cellcolor[rgb]{ .851,  .851,  .851}Ours & \cellcolor[rgb]{ .851,  .851,  .851}0.0000  & \cellcolor[rgb]{ .851,  .851,  .851}0.0000  & \cellcolor[rgb]{ .851,  .851,  .851}0.0000  & \cellcolor[rgb]{ .851,  .851,  .851}0.0004  \\
    \bottomrule
    \end{tabular}}%
  \label{tab:Size}%
\end{table}%

\begin{table}[!t]
  \centering
  \caption{The generality of EntroGuard in privacy-preserving capabilities when facing different attack models.}

  \resizebox{0.95\columnwidth}{!}{
    \begin{tabular}{clrrrr}
    \toprule
    \multicolumn{1}{c}{Attack model} & \multicolumn{1}{c}{Method} & \multicolumn{1}{c}{ROUGE} & \multicolumn{1}{c}{BLEU} & \multicolumn{1}{c}{EMR} & \multicolumn{1}{c}{BiNLI} \\
    \midrule
    \multirow{4}[2]{*}{GPT-2} & Original & 0.6427  & 0.6907  & 0.3377  & 0.7370  \\
          & \cellcolor[rgb]{ .851,  .851,  .851}GPT-2 & \cellcolor[rgb]{ .851,  .851,  .851}0.0000  & \cellcolor[rgb]{ .851,  .851,  .851}0.0000  & \cellcolor[rgb]{ .851,  .851,  .851}0.0000  & \cellcolor[rgb]{ .851,  .851,  .851}0.0008  \\
          & \cellcolor[rgb]{ .851,  .851,  .851}DialoGPT & \cellcolor[rgb]{ .851,  .851,  .851}0.0000  & \cellcolor[rgb]{ .851,  .851,  .851}0.0000  & \cellcolor[rgb]{ .851,  .851,  .851}0.0000  & \cellcolor[rgb]{ .851,  .851,  .851}0.0004  \\
          & \cellcolor[rgb]{ .851,  .851,  .851}Llama3 & \cellcolor[rgb]{ .851,  .851,  .851}0.0000  & \cellcolor[rgb]{ .851,  .851,  .851}0.0000  & \cellcolor[rgb]{ .851,  .851,  .851}0.0000  & \cellcolor[rgb]{ .851,  .851,  .851}0.0004  \\
    \midrule
    \multirow{4}[2]{*}{DialoGPT} & Original & 0.6142  & 0.6656  & 0.3001  & 0.7005  \\
          & \cellcolor[rgb]{ .851,  .851,  .851}GPT-2 & \cellcolor[rgb]{ .851,  .851,  .851}0.0000  & \cellcolor[rgb]{ .851,  .851,  .851}0.0000  & \cellcolor[rgb]{ .851,  .851,  .851}0.0000  & \cellcolor[rgb]{ .851,  .851,  .851}0.0004  \\
          & \cellcolor[rgb]{ .851,  .851,  .851}DialoGPT & \cellcolor[rgb]{ .851,  .851,  .851}0.0000  & \cellcolor[rgb]{ .851,  .851,  .851}0.0000  & \cellcolor[rgb]{ .851,  .851,  .851}0.0000  & \cellcolor[rgb]{ .851,  .851,  .851}0.0005  \\
          & \cellcolor[rgb]{ .851,  .851,  .851}Llama3 & \cellcolor[rgb]{ .851,  .851,  .851}0.0000  & \cellcolor[rgb]{ .851,  .851,  .851}0.0000  & \cellcolor[rgb]{ .851,  .851,  .851}0.0000  & \cellcolor[rgb]{ .851,  .851,  .851}0.0004  \\
    \midrule
    \multirow{4}[2]{*}{Llama3} & Original & 0.6813  & 0.7003  & 0.3929  & 0.7656  \\
          & \cellcolor[rgb]{ .851,  .851,  .851}GPT-2 & \cellcolor[rgb]{ .851,  .851,  .851}0.0142  & \cellcolor[rgb]{ .851,  .851,  .851}0.0256  & \cellcolor[rgb]{ .851,  .851,  .851}0.0001  & \cellcolor[rgb]{ .851,  .851,  .851}0.0121  \\
          & \cellcolor[rgb]{ .851,  .851,  .851}DialoGPT & \cellcolor[rgb]{ .851,  .851,  .851}0.0752  & \cellcolor[rgb]{ .851,  .851,  .851}0.0889  & \cellcolor[rgb]{ .851,  .851,  .851}0.0004  & \cellcolor[rgb]{ .851,  .851,  .851}0.0518  \\
          & \cellcolor[rgb]{ .851,  .851,  .851}Llama3 & \cellcolor[rgb]{ .851,  .851,  .851}0.0178  & \cellcolor[rgb]{ .851,  .851,  .851}0.0339  & \cellcolor[rgb]{ .851,  .851,  .851}0.0000  & \cellcolor[rgb]{ .851,  .851,  .851}0.0064  \\
    \bottomrule
    \end{tabular}}%
  \label{tab:Transferability}%
\end{table}%

\textbf{EntroGuard effectively mitigates privacy leakage at the logit level.} While exact-match and semantic metrics such as ROUGE and BiNLI effectively measure privacy leakage in the final decoded text, relying solely on top-1 text outputs may overlook underlying vulnerabilities. A model experiencing decoding collapse might output repetitive meaningless tokens such as $\langle\text{endoftext}\rangle$, masking the fact that actual sensitive tokens could still rank highly in the underlying probability distribution, a phenomenon known as logit leakage. To rigorously verify that EntroGuard fundamentally scrubs sensitive information rather than merely disrupting the decoding process, we conduct a comprehensive logit-level analysis of the attacker's autoregressive generation. We evaluate three metric categories, namely \textit{Top-$k$ Presence} (i.e., Token Recall@$k$) to represent the fraction of ground-truth sensitive tokens appearing in the top-$k$ predictions across any decoding step, \textit{Probability Mass} to indicate the softmax probability assigned to sensitive tokens, and \textit{Rank Metrics} to denote the ranking of sensitive tokens in the predicted distribution. As shown in \autoref{tab:logit_leakage}, the protection is clearly not an artifact of decoding collapse. The Token Recall@10 is merely 0.14\%, indicating that sensitive tokens almost never appear in the top-10 candidates. Furthermore, the mean rank of a sensitive token during generation drops to 12641.71, with a negligible average probability of 0.000158. These metrics demonstrate that EntroGuard successfully submerges sensitive tokens within the vast background vocabulary, rendering them statistically indistinguishable from other irrelevant candidates and effectively mitigating logit-level leakage risks.

\textbf{EntroGuard effectively withstands different decoding algorithms.} To verify that the privacy improvements remain robust under different generation configurations, we evaluate the attacker utilizing various decoding strategies. We test greedy decoding alongside beam search with varying widths such as $B=5$, $10$, and $20$. As shown in \autoref{tab:decoding_strategies}, EntroGuard consistently maintains high privacy-preserving capabilities. Even when the attacker employs aggressive search strategies to extensively explore the output space, text-level leakage metrics including ROUGE, BLEU, and EMR remain at zero. Concurrently, the semantic leakage metric BiNLI remains negligible across all configurations. These results demonstrate that our defense successfully withstands even the most exhaustive decoding search algorithms.

\begin{table}[!t]
  \centering
  \caption{Privacy leakage under different decoding strategies.}
  \resizebox{0.95\columnwidth}{!}{
    \begin{tabular}{lrrrr}
    \toprule
    \multicolumn{1}{c}{Decoding Strategy} & \multicolumn{1}{c}{ROUGE} & \multicolumn{1}{c}{BLEU} & \multicolumn{1}{c}{EMR} & \multicolumn{1}{c}{BiNLI} \\
    \midrule
    Greedy Decoding       & 0.0000 & 0.0000 & 0.0000 & 0.0004 \\
    Beam Search ($B=5$)   & 0.0000 & 0.0000 & 0.0000 & 0.0008 \\
    Beam Search ($B=10$)  & 0.0000 & 0.0000 & 0.0000 & 0.0008 \\
    Beam Search ($B=20$)  & 0.0000 & 0.0000 & 0.0000 & 0.0007 \\
    \bottomrule
    \end{tabular}}%
  \label{tab:decoding_strategies}%
\end{table}

\textbf{EntroGuard can accommodate different embedding models of different sizes.} Given that the size of embedding models deployed on different end devices may be heterogeneous, it is crucial to evaluate the applicability of our scheme across models with different parameter scales. To this end, we evaluate privacy-preserving capability of our approach utilizing Sentence-T5 models of varying sizes, including sentence-t5-base, sentence-t5-large, sentence-t5-xl, and sentence-t5-xxl. As shown in \autoref{tab:Size}, our scheme consistently reduces ROUGE, BLEU, EMR, and BiNLI scores to zero across all model sizes, demonstrating its robust protection capability regardless of model scale.

\begin{table}[!t]
  \centering
  \caption{Performance metrics for different values of $\lambda$}
  \resizebox{0.95\columnwidth}{!}{
    \begin{tabular}{cccccc}
    \toprule
    \multicolumn{1}{c}{$\lambda$} & \multicolumn{1}{c}{NDCG} & \multicolumn{1}{c}{MAP} & \multicolumn{1}{c}{Precision} & \multicolumn{1}{c}{BLEU} & \multicolumn{1}{c}{BiNLI} \\
    \midrule
    0.15  & 0.4923  & 0.0472  & 0.5861  & 0.0000  & 0.0008 \\
    0.35  & 0.4931  & 0.0473  & 0.5814  & 0.0000  & 0.0008 \\
    0.55  & 0.5020  & 0.0477  & 0.5861  & 0.0000  & 0.0006 \\
    0.75  & 0.4921  & 0.0471  & 0.5767  & 0.0000  & 0.0007 \\
    0.95  & 0.4933  & 0.0478  & 0.5954  & 0.0000  & 0.0008 \\
    \bottomrule
    \end{tabular}}%
  \label{tab:lambda_metrics}%
\end{table}%

\textbf{EntroGuard has good generality to different attack models.}
To evaluate the robustness of our scheme against unknown attack models, we evaluated its privacy-preserving effectiveness using GPT-2, DialoGPT, and Llama3 as attack models, while training EntroGuard on a separate surrogate model. As shown in \autoref{tab:Transferability}, our scheme also consistently reduces ROUGE, BLEU, EMR, and BiNLI scores to near-zero values across all attack models. Notably, even confronting a relatively stronger attacker (e.g., Llama3) with a relatively weaker surrogate model (e.g., DialoGPT), the degree of textual privacy leakage is still low (e.g., BLEU 0.0889, BiNLI 0.0518), indicating that our scheme has strong privacy-preserving ability with high generality against different attack models.

\begin{figure}[!t]
\centering
\subfloat[Inference with CPUs]{%
    \includegraphics[width=0.8\columnwidth]{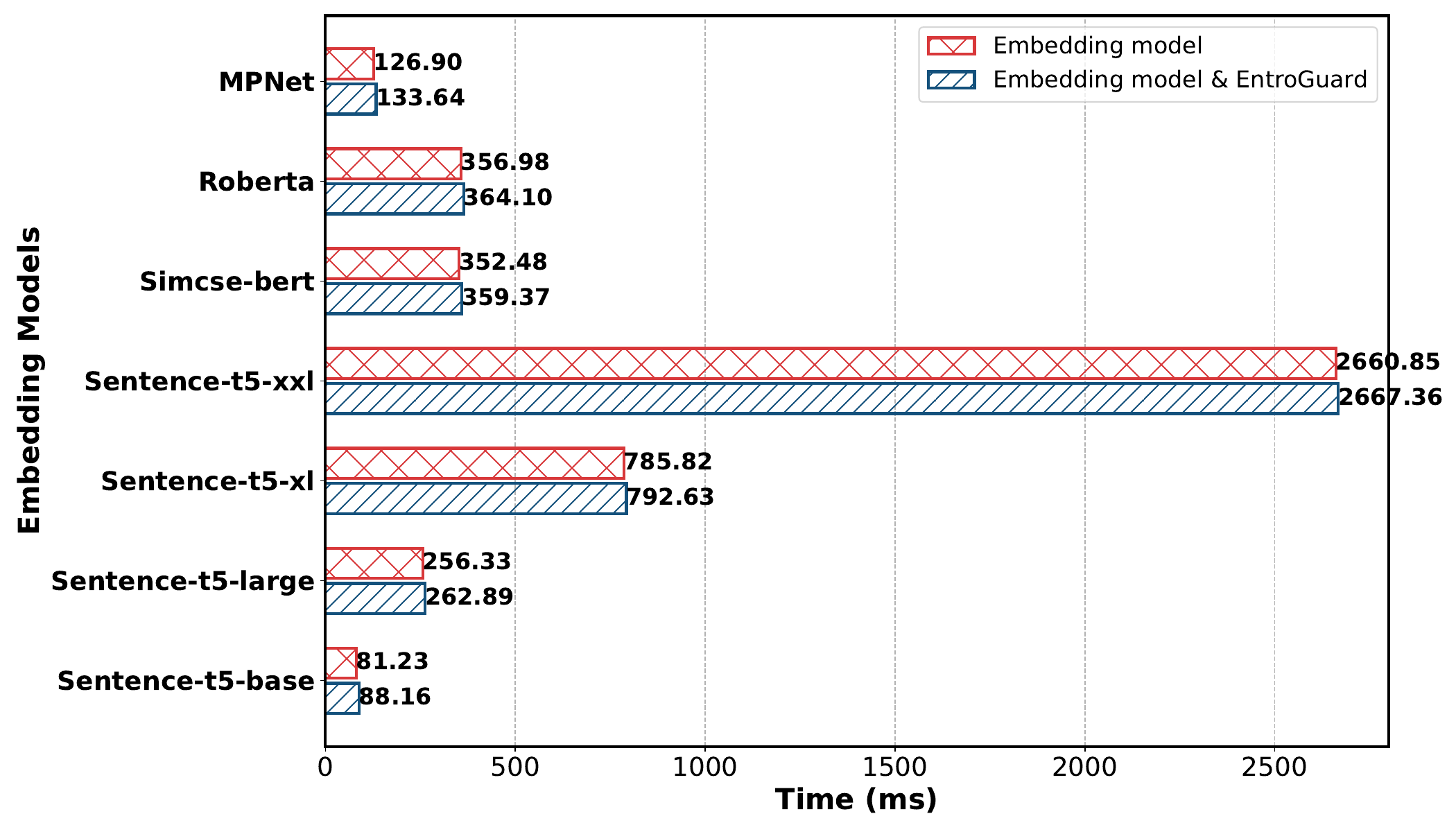}

    \label{fig:end_device_cpu}
}
\hfill
\subfloat[Inference with GPUs]{%
    \includegraphics[width=0.8\columnwidth]{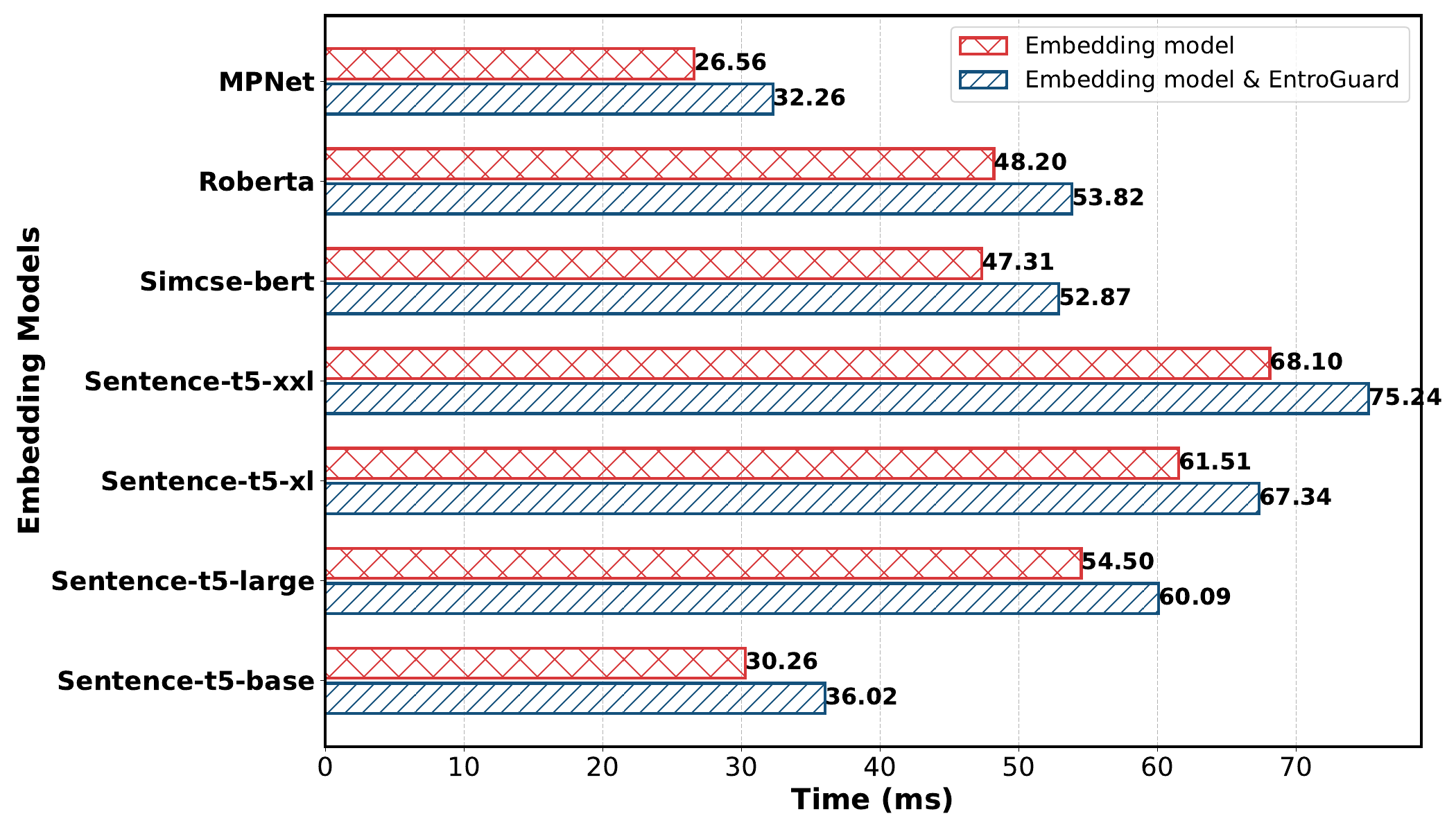}
    \label{fig:end_device_gpu}
}

\caption{Comparison of inference efficiency on end devices using different hardware.}
\label{fig:end_device_comparison}

\end{figure}

\textbf{Sensitivity Analysis of Scaling Factor $\lambda$.} The scaling factor $\lambda$ controls the adjustment step size in BPA. \autoref{tab:lambda_metrics} shows that privacy remains consistently strong across all $\lambda$, with BLEU and BiNLI scores near zero. Despite this practical stability, we optimize $\lambda$ for its theoretical noise retention. A smaller $\lambda$ (e.g., 0.15) may cause an excessive reduction of perturbation intensity, which theoretically weakens privacy protection. Conversely, $\lambda=0.95$ enables finer adjustments, keeping perturbations closer to the upper retrieval bound to maximize attacker disruption. Although a larger $\lambda$ requires more adaptation steps, the overhead is negligible, i.e., 0.016 ms per operation on a single NVIDIA RTX 5880 GPU. Thus, we set $\lambda=0.95$ as the default to achieve a favorable empirical privacy-utility trade-off without noticeable delay.

\begin{table}[!t]
  \centering
  \caption{Ablation study on the loss components of EntroGuard.}
  \setlength{\tabcolsep}{3pt}
  \resizebox{0.95\columnwidth}{!}{
    \begin{tabular}{lccccccc}
    \toprule
    \multicolumn{1}{c}{Method} & \multicolumn{1}{c}{NDCG} & \multicolumn{1}{c}{MAP} & \multicolumn{1}{c}{Precision} & \multicolumn{1}{c}{ROUGE} & \multicolumn{1}{c}{BLEU} & \multicolumn{1}{c}{EMR} & \multicolumn{1}{c}{BiNLI} \\
    \midrule
    w/o $L_{Entropy}$  & 0.5046 & 0.0482 & 0.5907 & 0.0155 & 0.0123 & 0.0011 & 0.0065 \\
    w/o $L_{CE}$       & 0.5047 & 0.0481 & 0.5954 & 0.0114 & 0.0148 & 0.0002 & 0.0023 \\
    Full Loss          & 0.4933 & 0.0478 & 0.5954 & 0.0000 & 0.0000 & 0.0000 & 0.0008 \\
    \bottomrule
    \end{tabular}}%
  \label{tab:ablation_loss}%
\end{table}

\textbf{EntroGuard effectively protects privacy by combining both loss components.} To evaluate how each component contributes and verify that our method does not simply overfit the GEIA training objective, we conduct an ablation study. Specifically, we train the perturbation generator under three configurations including $L_{Similarity}+L_{CE}$, $L_{Similarity}+L_{Entropy}$, and the full loss. The $L_{Similarity}$ term is included in all settings to maintain retrieval utility. As shown in \autoref{tab:ablation_loss}, results on the MS MARCO dataset demonstrate the necessity of both privacy components. The setting using only $L_{Entropy}$ effectively disturbs the recovery process from the intermediate blocks to reduce semantic leakage, yielding a low BiNLI of 0.0023. Meanwhile, using $L_{CE}$ alone provides privacy protection, though slightly less effective. By integrating both mechanisms, the full loss prevents privacy leakage while maintaining high retrieval performance.

\begin{table}[!t]
  \centering
  \caption{Storage and computational overhead of the EntroGuard}
  \setlength{\tabcolsep}{3pt}
  \resizebox{0.95\columnwidth}{!}{
    \begin{tabular}{lcccccc}
    \toprule
    \multirow{2}{*}{\begin{tabular}[c]{@{}l@{}}Embedding \\ model\end{tabular}} & \multicolumn{3}{c}{Storage size (MB)} & \multicolumn{3}{c}{Complexity (MACs)} \\
    \cmidrule(lr){2-4} \cmidrule(lr){5-7}
          & Model & EntroGuard & Overhead & Model & EntroGuard & Overhead \\
    \midrule
    Base  & 418.20   & 11.34 & \textbf{2.71\%}  & 3.397 G   & 2.976 M & \textbf{0.088\%} \\
    Large & 1277.69  & 11.34 & \textbf{0.89\%}  & 12.080 G  & 2.976 M & \textbf{0.025\%} \\
    XL    & 4733.70  & 11.34 & \textbf{0.24\%}  & 48.318 G  & 2.976 M & \textbf{0.006\%} \\
    XXL   & 18557.71 & 11.34 & \textbf{0.06\%}  & 193.274 G & 2.976 M & \textbf{0.002\%} \\
    \bottomrule
    \end{tabular}}%
  \label{tab:entroguard_overhead}%
\end{table}

\textbf{Stability Across Random Seeds.} We further evaluate EntroGuard on MS MARCO with Sentence-T5 as the embedding model across five random seeds (256, 512, 1024, 2048, and 4096). The results show negligible variance in both privacy leakage and retrieval performance, with standard deviations of $2.56 \times 10^{-8}$ for BiNLI and $6.24 \times 10^{-6}$ for NDCG. These results indicate stable privacy protection and retrieval preservation under this representative setting.

\subsection{Performance Overhead and Edge Deployment} \label{sec:Experiments4}
\begin{figure}[!t]
\centering
\includegraphics[width=0.8\columnwidth]{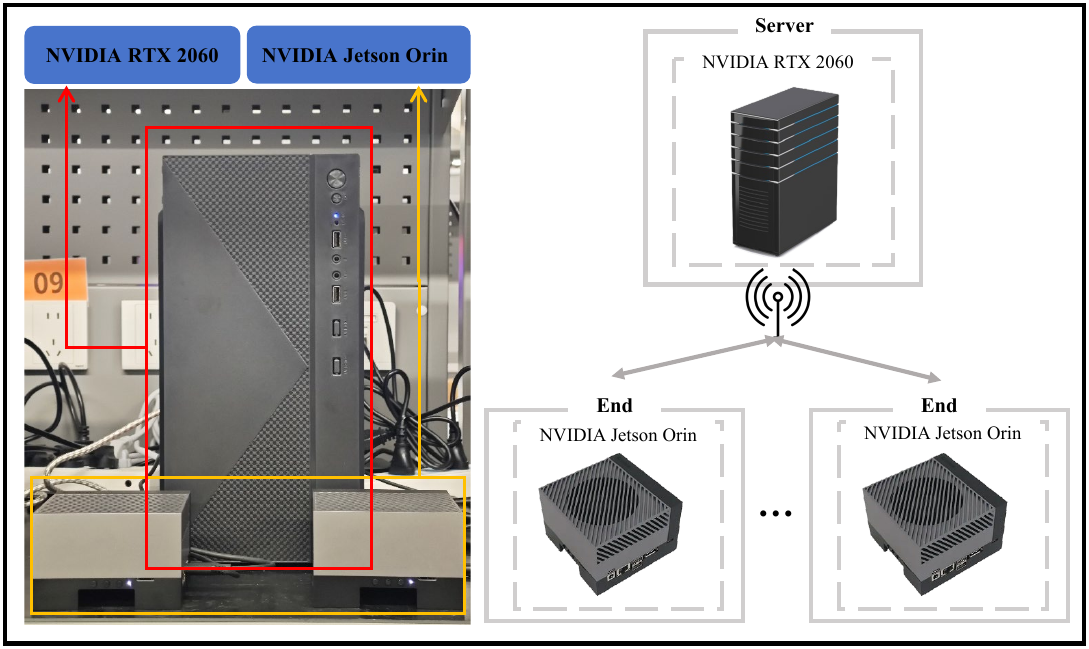}

\caption{Testbed in real world.}
\label{fig:end_device}

\end{figure}

In this section, we first evaluate the offline training time, then analyze the theoretical complexity and footprint, and finally evaluate the efficiency of our proposed scheme on real-world end devices as shown in \autoref{fig:end_device}.

\textbf{EntroGuard incurs a modest one-time offline training overhead.} Before discussing the on-device performance, we first quantify the offline preparation phase. Training the surrogate model and optimizing the EntroGuard perturbation generator on the MS MARCO dataset requires approximately 5 hours and 200 minutes respectively, utilizing a single NVIDIA RTX 5880 GPU. Notably, these represent a one-time offline setup. Once trained, this process yields a lightweight generator ready for edge deployment.

\textbf{EntroGuard possesses a lightweight theoretical complexity and footprint.} To demonstrate the universal viability of the resulting EntroGuard generator across diverse mobile devices, we provide a fundamental theoretical complexity analysis as detailed in \autoref{tab:entroguard_overhead}. The generator maintains a constant and ultra-lightweight footprint, requiring 11.34 MB of static storage for its parameters and 2.976 million MACs per inference. Compared to the base embedding models such as the Sentence-T5 series, the maximum storage overhead is only 2.71\%, and the computational overhead is less than 0.088\%. This negligible theoretical cost ensures that EntroGuard is perfectly viable for severely resource-constrained environments.

\textbf{EntroGuard can be effectively integrated into end devices and achieve efficient inference.} Guided by the aforementioned theoretical analysis, we have implemented EntroGuard in real-world end devices, i.e., the Jetson AGX Orin Developer Kit~\cite{jetson}, where seven embedded models with different structures and different parameter sizes are deployed. Specifically, EntroGuard can be deployed as a plug-in, without needing to know the detailed structure of the embedding model. The actual storage footprint for the parameters is slightly increased to approximately 13MB due to serialization overhead. In practice, the output of the original embedding model is directly fed into EntroGuard for processing, which then generates the privacy-preserving embeddings. To evaluate EntroGuard's efficiency, we measured the computation time on both CPUs and GPUs by averaging the time over 17K sequential inputs. For CPU inference, the time cost increases by only 3\% compared to the original embedding models when EntroGuard is applied. More detailed time costs on CPUs are shown in \cref{fig:end_device_cpu}. For GPU inference, compared to the original models, the time cost increases by 13\% approximately when EntroGuard is plugged in, whose detailed time costs on GPUs are provided in \cref{fig:end_device_gpu}. Notably, the absolute time to run our scheme on both CPUs and GPUs is relatively short, i.e., approximately 7 ms on CPUs and 5 ms on GPUs. Since the inference time of the embedding model on GPUs reduces, the relative percentage increase in time consumption is more noticeable. In summary, EntroGuard can be easily integrated into existing black-box embedding models, providing efficient protection for embeddings with an acceptable impact on processing time.

\begin{table}[!t]
  \centering
  \caption{Performance of EntroGuard after 10 rounds of iterative defense updating.}
  \resizebox{0.95\columnwidth}{!}{
    \begin{tabular}{cccccccc}
    \toprule
    \multicolumn{1}{c}{Method} & \multicolumn{1}{c}{NDCG} & \multicolumn{1}{c}{MAP} & \multicolumn{1}{c}{Precision} & \multicolumn{1}{c}{ROUGE} & \multicolumn{1}{c}{BLEU} & \multicolumn{1}{c}{EMR} & \multicolumn{1}{c}{BiNLI} \\
    \midrule
    EntroGuard & 0.4855 & 0.0481 & 0.5814 & 0.0104 & 0.0122 & 0.0000 & 0.0013 \\
    \bottomrule
    \end{tabular}}%
  \label{tab_adaptive_attack2}%
\end{table}

\subsection{Discussion}
\textbf{Iterative Defense Updating.} In real-world deployments, attackers may continuously update recovery models to bypass existing defenses. While acquiring massive training datasets of private texts paired with protected embeddings is a major obstacle for adversaries, we evaluate whether EntroGuard's decoupled plug-in architecture can be efficiently updated in response to such evolving attackers. We construct a multi-round iterative defense-updating experiment. In each round, the adversary first retrains its recovery model on the protected embeddings generated by the current $G_\theta$, after which we update $G_\theta$ through lightweight retraining without modifying the base embedding model. As shown in \autoref{tab_adaptive_attack2}, following the defense update in the tenth round, the BiNLI and BLEU scores are reduced to 0.0013 and 0.0122, respectively, while the NDCG and Precision remain at 0.4855 and 0.5814. These results show that EntroGuard can be efficiently updated on demand to defend against evolving threats.

\textbf{Architectural Scope and Future Directions.}
Our current implementation and evaluation mainly focus on Transformer-based recovery models, which are commonly adopted by existing powerful learning-based EIAs. As discussed in \cref{sec:Experiments5}, EntroGuard shows transferability across different Transformer-based attackers, but its effectiveness against non-Transformer recovery architectures has not been empirically evaluated. Exploring whether and how entropy-driven protection can be extended to emerging non-Transformer recovery architectures remains an important direction for future work.

\section{Conclusions}
In this paper, we introduce EntroGuard, a novel privacy-preserving transmission approach designed to protect the privacy of text embeddings transmitted from end devices while maintaining retrieval accuracy in cloud databases without requiring additional cooperation from the cloud. Meanwhile, EntroGuard can be efficiently integrated into the existing embedding model on end devices with acceptable overhead. Furthermore, we propose BiNLI, a metric for evaluating sentence privacy based on semantic similarity, which enables a more comprehensive quantification of text privacy leakage. Extensive experiments demonstrate that EntroGuard outperforms existing text embedding protection methods, offering superior privacy protection with minimal loss in retrieval performance.

\bibliographystyle{IEEEtran}
\bibliography{sample-base}

\end{document}